\documentclass{emulateapj}

\newcommand{\Msun}{ M_{\odot}}
\newcommand{\Mdot}{\dot{M}}
\newcommand{\Mbh}{M}
\newcommand{\teff}{T_{\rm eff}}

\begin{document}
\title{The Radiative Efficiency of Accretion Flows in Individual AGN}
\author{Shane W. Davis\altaffilmark{1,2} and Ari Laor\altaffilmark{3}}
\altaffiltext{1}{Institute for Advanced Study, Einstein Drive, Princeton, NJ 08540}
\altaffiltext{2}{Canadian Institute for Theoretical Astrophysics. Toronto, ON M5S3H4, Canada}
\altaffiltext{3}{Physics Department, Technion, Haifa~32000, Israel}

\begin{abstract}
  The radiative efficiency of AGN is commonly estimated based on the
  total mass accreted and the total AGN light emitted per unit volume
  in the universe integrated over time (the Soltan argument). In
  individual AGN, thin accretion disk model spectral fits can be used
  to deduce the absolute accretion rate $\Mdot$, if the black hole
  mass $\Mbh$ is known. The radiative efficiency $\eta$ is then set by
  the ratio of the bolometric luminosity $L_{\rm bol}$ to $\Mdot c^2$.
  We apply this method to determine $\eta$ in a sample of 80 PG
  quasars with well determined $L_{\rm bol}$, where $\Mdot$ is set by
  thin accretion disk model fits to the optical luminosity density,
  and the $\Mbh$ determination based on the bulge stellar velocity
  dispersion (13 objects) or the broad line region (BLR). For the
  BLR-based masses, we derive a mean $\log \eta=-1.05\pm 0.52$
  consistent with the Soltan argument based estimates.  We find a
  strong correlation of $\eta$ with $\Mbh$, rising from $\eta\sim
  0.03$ at $\Mbh=10^7\Msun$ and $L/L_{\rm Edd}\sim 1$ to $\eta\sim
  0.4$ at $\Mbh=10^9\Msun$ and $L/L_{\rm Edd}\sim 0.3$.  This trend is
  related to the overall uniformity of $L_{\rm opt}/L_{\rm bol}$ in
  our sample, particularly the lack of the expected increase in
  $L_{\rm opt}/L_{\rm bol}$ with increasing $\Mbh$ (and decreasing
  $L/L_{\rm Edd}$), which is a generic property of thermal disk
  emission at fixed $\eta$. The significant uncertainty in the $\Mbh$
  determination is not large enough to remove the correlation. 
  The rising $\eta$ with $\Mbh$ may imply a rise in
  the black hole spin with $\Mbh$, as proposed based on other indirect
  arguments.

\end{abstract}
\keywords{accretion, accretion disks --- black hole physics --- galaxies: active --- galaxies: quasars: general}

\section{Introduction}

Material falling in nearly circular orbits onto a black hole
(hereafter, BH) looses a fraction of its rest mass energy during the
infall. The lost energy is emitted as an outflow of radiation and
particles (and potentially Poynting flux). A measurement of the
fraction of mass inflow $\Mdot$ converted to radiation $L_{\rm bol}$,
provides a measure of the radiation efficiency $\eta\equiv L_{\rm
  bol}/\Mdot c^2$.

In the ``standard'' accretion disk (AD) model
\citep{ss73,nt73}, the BH spin $a_*$ determines $\eta$ because it sets
the marginally stable orbit, $r_{\rm ms}$, beyond which the material
is assumed to fall into the BH without loosing further energy.  Since
the total efficiency is a rising monotonic function of $a_*$, the
measured $\eta$ provides a lower limit on $a_*$.  The value of $a_*$
is important because it tells us how the BH mass $\Mbh$ grew.  If it
grew mostly through a single event (major merger or continuous gas
accretion) then $a_*$ will be close to unity. If it grew through a
series of independent events (minor mergers, episodic accretions),
then $a_*$ will be close to zero \citep{hb03,gsm04,vol05,kp06,bv08}.

In the absence of torques near $r_{\rm ms}$, the value of $a_*$ sets,
through the value of $r_{\rm ms}$, the spectral energy distribution of
the accretion disk \citep{cun75,ks84,sm89,ln89,sk98,hub00}, the
rotation of the polarization angle of the AD emission,
\citep{csp80,lnp90,dov08,sk09}, and the profile of lines emitted by
the AD \citep{fab89,koj91,lao91,dab97,bd04,br06,rf08}.  These methods
are currently limited by the available quality of the data, and by
potential uncertainties in our models of the AD structure. As a
result, we do not yet have precision measurements of $a_*$ in more
than a few objects. Thus, an additional independent constraint on
$a_*$ based on $\eta$, is useful.

A determination of $\eta$ can potentially provide an upper limit on
the additional power which may be generated by the accretion in a
jet/wind outflow.  Without torques, there is an upper limit on the
total efficiency of 40\% for $a_*=1$, or 31\%, for the maximal spin
within an AD of $a_*=0.998$ \citep{tho74}. Such outflows are important
as they can couple to the surrounding gas more efficiently than
radiation, and may significantly affect the host galaxy evolution
(e.g.  \citealt{mn07}), suppress cluster cooling flows (e.g.
\citealt{chu02,all06}), and may be relevant to the correlation of the
black hole mass with the galaxy properties \citep[ and
  citations thereafter]{mag98}. The implied jet power of AGN in
cooling flow clusters can be significantly larger or smaller than the
radiative power, depending on the AGN luminosity (e.g.
\citealt{sha08b,mh08,cb09}). Clearly, it is useful to get an
independent upper limit on the ratio of mechanical/radiative power,
based on a direct determination of $\eta$.  If magnetohydrodynamics
torques \citep{gam99,kro99a,mg04,dhk03} are present, then the maximum
efficiency can (instantaneously) exceed the limits for a no-torque disk
and even exceed unity (see e.g. \citealt{ak00}) as the flow taps the
spin energy of the BH.  Therefore, credible estimates of such large
efficiencies would provide evidence that such torques are present in
real accretion flows.

\citet{sol82} noted that the global AGN average radiative efficiency,
$\eta_{\rm av}$, can be estimated for the AGN population by comparing
the integrated $\Mbh$ per unit volume at the current epoch, with the
integrated AGN luminosity per unit volume over time. \citet{sol82}
also showed that $\eta_{\rm av}$ is elegantly independent of the
cosmological model (a major unknown at that time). Recent studies
based on the Soltan argument lead to $\eta_{\rm av}\ga 0.1$ (e.g.
\citealt{yt02,erz02,mar04,bar05}).  This method has also been used to
estimate the time and luminosity dependence of $\eta_{\rm av}$ through
more detailed modeling (e.g. \citealt{hnh06,swm09,wan09,rf09}), but
the derived values are significantly uncertain.

The purpose of this paper is to discuss a method to derive $\eta$
directly in individual AGN. The method assumes that the optically
emitting regions of QSOs are accretion powered and radiatively
efficient, thus gravitational binding energy is dissipated and
radiated locally within the AD. The corresponding thin AD models were
calculated to increasing levels of details, from the simple local
blackbody approximation to stellar atmosphere like models where the
vertical structure and the local spectrum are calculated with
increasing accuracy (see \citealt{hub00} and references therein).  The
integrated thin AD luminosity density $L_{\nu}$ turns out to be
largely set by $\Mdot$ and $\Mbh$. Thus, one can derive $\Mdot$ based
on the observed $L_{\nu}$, if $\Mbh$ is known.  This method has been
used previously by \citet{col02} and \citet{bz03}, using simple
analytic expressions, valid at long wavelengths for the emission of an
Newtonian, thin, blackbody AD (e.g. \citealt{bec87}), to
determine $\Mdot$ in a sample of AGN.  \citet{col02} assumed a value
of $\eta$ to estimate $L_{\rm bol}$ and inferred that many low $\Mbh$
AGN must be super-Eddington accretors.  \citet{bz03} estimated $L_{\rm
  bol}$ independently for each object, which they then used to
estimate $\eta$, yielding an average $\log~\eta=-1.77\pm 0.49$ in a
sample of radio-quiet AGN, and $\log~\eta=-0.90\pm 0.62$ in radio-loud
AGN.

Observations indicate that simple thin AD model cannot reproduce the overall
SED. This is due to reprocessing (IR), Comptonization in a corona
(X-ray), radiative transfer effects in the inner AD, a thick AD, etc'.
Our method relies on the viability of the simple thin disk
approximation in the relative outer parts of the AD, which dominate
the optical emission. The above effects are likely insignificant in
this outer part of the AD. Thus, the redistribution of the AD radiation by
these effects will not affect the measurement of $\eta$, as long as we
measure the total SED, irrespective of its exact production mechanism.

Here we derive $\Mdot$ based on relatively sophisticated AD models,
which include relativistic effects on the disk structure and photon
propagation to the observer, and solve simultaneously for the vertical
structure and radiative transfer of the disk. We apply the method to
the PG quasar sample \citep{sg83}, where $L_{\rm bol}$ is estimated
based on high quality optical \citep{neu87}, UV \citep{bl05}, far UV
\citep{sco04}, and soft X-ray \citep{blw00} observations, and $\Mbh$
is derived based on high quality spectroscopy of the H$\beta$ region
by \citet{bg92}. The paper is organized as follows, in \S 2 we review
the simple analytic derivation of $\Mdot$, and demonstrate that the AD
$L_{\nu}$ in the optical regime is rather well determined by the local
blackbody AD models, and is only slightly modified by taking into
account the vertical disk structure.  We also show that the optical
$L_{\nu}$ is only weakly dependent on the radial disk structure, as
set by $a_*$. We then derive $\Mdot$ for our sample.  In \S 3 we
estimate $L_{\rm bol}$, and combined with $\Mdot$ use it to compute
$\eta$.  We discuss the correlation, or lack thereof, of $\eta$ with
parameters of interest, particularly $\Mbh$.  In \S 4 we discuss
various systematic effects which can affect the value of $\eta$ and
the observed correlation, in particular the uncertainty in $\Mbh$,
disk inclination, optical thickness of the AD emission,
self-illumination, foreground extinction, and mass outflows. We
summarize our conclusions in \S 5.

\section{Estimating the Accretion Rate}
\label{mdot}

\subsection{Mass Estimates}
\label{mass}

The spectral based methods for computing $\Mdot$ outlined in \S
\ref{simple} and \ref{model} require $\Mbh$ estimates.  We consider
two sets of estimates, $M_{\rm BLR}$ and $M_\sigma$, based on the
broad emission line widths and the $\Mbh - \sigma_*$ correlation,
respectively.

The first method requires characteristic velocities and radii for the
Broad Line Region (BLR).  We use the luminosity radius relation of
\citet{kas05} to compute the BLR radius $R_{\rm BLR}$, but with
$L_{\rm opt}=\nu L_\nu$ measured at 4861\AA~instead of 5100\AA.
Inserting this relation into equation (5) of \citet{kas00} and using
the H$\beta$ FWHM $v$ \citep{bg92}, we can compute
\begin{equation}
M_{\rm BLR}=1.5\times 10^8 \Msun L_{\rm opt,45}^{0.69}  v_{3000}^2,
\label{eq:mass}
\end{equation}
where $L_{\rm opt,45}$ is $L_{\rm opt}/10^{45} \rm \; erg \; s^{-1}$,
and $v_{3000} = {\rm H}\beta$ FWHM$/(3000$ $\rm \; km \; s^{-1})$.
The quantities $L_{\rm opt,45}, v_{3000}$, and $M_{\rm BLR}$ are
reported in Table~1.

A second method relies on the tight correlation between $\Mbh$ and the
stellar velocity dispersion $\sigma_*$ \citep{fm00,geb00}.  The major
difficulty with this application is that the quasar light dwarfs the
emission from the rest of the galaxy, making it particularly
challenging to measure $\sigma_*$.  Therefore, $\sigma_*$ is only
available for a handful of sources, and the measurements are likely
less robust than those of inactive galaxies.  For 13 of our sources,
we also use $\sigma_*$ estimates from \citet{das07} and \citet{ws08}
to compute $M_\sigma$ with the \citet{tre02} relation, and these
values are reported in the second column of Table~2.

\subsection{Analytic Method}
\label{simple}

The flux per unit area emitted by a thin AD is \citep{ss73}
\begin{equation}
 F=\frac{3}{8\pi}\frac{G\Mdot \Mbh}{R^3}f_c(r,a)
\end{equation}
where $R$ is the radius, and $f_c(r,a)$ is a dimensionless factor,
typically of order unity, which takes into account both the no-torque
inner boundary condition, and the relativistic effects
\citep{nt73,pt74,rh95}. This factor, which approaches unity when $r\gg
1$, depends on $a_*$ and on the dimensionless radius $r\equiv R/R_g$
where $R_g\equiv G\Mbh/c^2$.

We assume the disk emits locally as a blackbody at the effective
temperature $\teff\equiv (F/\sigma)^{1/4}$. The full disk
spectrum can now be obtained by integrating over the disk surface
\begin{equation}
L_\nu=8 \pi^2 \cos i R_g^2 \int^{r_{\rm out}}_{r_{\rm in}} 
B_\nu(\teff(r)) r dr,
\label{eq:lnu1}
\end{equation}
where $L_\nu$ is the observed specific intensity assumed to be
emitted over $4 \pi$
steradians, $i$ is the inclination to the line-of-sight, and
$B_\nu$ is the Planck function.

Since we are also primarily interested in the emission at relatively
large radius, we will ignore the $f_c(r,a)$ dependence and use
$\teff = T_0 r^{-3/4}$ with
\begin{equation}
T_0 \equiv \left(\frac{3 c^6}{8\pi G^2 \sigma} \right)^{1/4}
\frac{\Mdot^{1/4}}{\Mbh^{1/2}}.\label{eq:t0}
\end{equation}  
Defining $x\equiv h\nu/(k_{\rm B}
\teff)$ and switching integration variables yields
\begin{equation}
L_\nu = \frac{160}{\pi^3}
\left( \frac{6\pi^2 h G^2 c^2}{5} \right)^{1/3} \Theta \nu^{1/3}
\Mdot^{2/3} \Mbh^{2/3}\cos i ,
\label{eq:lnu2}
\end{equation}
where
\begin{equation}
\Theta(\nu,r_{\rm in},r_{\rm out}) \equiv
\int^{x_{\rm out}}_{x_{\rm in}} dx \frac{x^{8/3-1}}{\exp(x) - 1}.
\end{equation}
In the following, we will assume a constant $\Theta \approx 1.93$.
This is approximately correct when $\nu_{\rm in}\gg \nu\gg \nu_{\rm
  out}$, where $\nu_{\rm in,out}$ are the frequencies of the peak of
the emission at $r_{\rm in,out}$.  For most of the PG quasars, this is
a reasonable approximation as long as we evaluate $L_\nu$ in the
visible range, since $\nu_{\rm out}$ is expected to be in the IR,
while $\nu_{\rm in}$ typically in the EUV.

Defining $L_{\rm opt} \equiv \nu L_\nu$ at 4861 \AA~ and solving
equation (\ref{eq:lnu2}) for $\Mdot$, we find
\begin{equation}
\Mdot = 2.6 \Msun~{\rm yr}^{-1} \left( \frac{L_{\rm opt,45}}{\cos i}\right)^{3/2}
M_8^{-1},
\label{eq:mdot}
\end{equation}
where $M_8=\Mbh/10^8$ and $L_{\rm opt,45}=L_{\rm opt}/{\rm
  10^{45} \; erg \; s^{-1}}$ Thus, the absolute accretion rate, $\Mdot$, can
be estimated from the observed $L_{\rm opt}$, if $\Mbh$ and $i$ are known.

\begin{figure}[h]
\includegraphics*[width=\columnwidth]{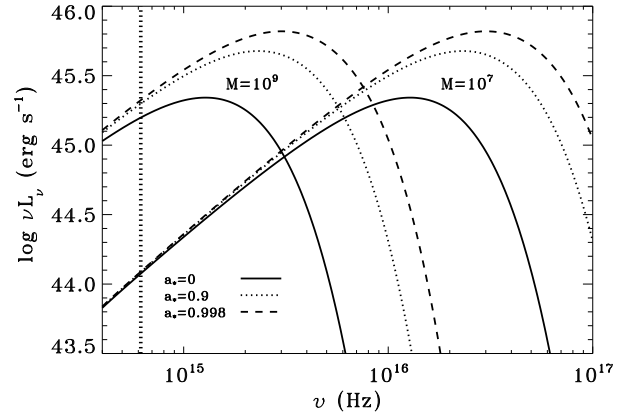}   
\caption{
Comparison of SEDs for local blackbody AD models, with different
values of $a_*$ and $M$. All models have $\Mdot=1 \Msun \rm \;
yr^{-1}$ and $\cos i=0.8$.  Note that the disk becomes hotter with
increasing $a_*$, but $L_{\rm opt}$ (at the vertical dashed line) is
generally independent of $a_*$ (excluding the $a_*=0$ $M=10^9$ model,
which is too cold to be a viable AGN SED anyhow).  Higher $M$ AD
models are colder, for a fixed $\Mdot$, as the SED $\nu_{\rm
  peak}\propto M^{-1/2}$, and are optically more luminous, as $L_{\rm
  opt}\propto M^{2/3}$. Thus, if $M$ is known, then $\Mdot$ is set by
$L_{\rm opt}$.
\label{f:spin_comp}}
\end{figure}

\subsection{Modeling Method}
\label{model}

In the previous section, we computed analytic expressions for $\Mdot$
from a simple, local blackbody model of an AD using several
approximations.  We now outline a calculation which takes advantage of
more sophisticated spectral models.  We use KERRTRANS
(\citealt{ago97}; see also \citealt{da09}) to calculate the disk
integrated spectrum from a fully relativistic disk model \citep{nt73}.
In this model the effects of varying the BH spin are included, with
the assumption made that the innermost radius of the disk corresponds
to the radius of marginal stability for circular orbits $r_{\rm ms}$,
and that no torque is present at this radius.  We further assume no
emission from inside $r_{\rm ms}$.

We consider models in which the local emission at the disk surface is
calculated with two different methods.  It is either assumed to be a
blackbody or it is computed using stellar-atmosphere-like calculations
of the disk vertical structure.  The latter (hereafter referred to as
TLUSTY models) are essentially equivalent to the models described in
\citet{hub00}\footnote{The physical content of the models is
  identical.  The only difference is that the spectrum for each
  annulus is computed via interpolation on a precomputed table of
  annuli, as described in \cite{dh06}.} and are computed using the
TLUSTY code \citep{hl95}.

The relativistic model has four parameters: $\Mbh$, $a_*$, $\Mdot$ and
$i$.  The spectra based on the full vertical structure calculations
also require a choice of $\alpha_{\rm SS}$ \citep{ss73} to determine
the disk surface density.  We assume $\alpha_{\rm SS}=0.01$ for all
models unless otherwise specified.

These parameters completely determine the emission at all frequencies,
and can be compared directly with the observations.  Specifically, we
will focus on matching to the observed luminosity at 4861\AA.  We use
the mass estimates discussed in \S\ref{mass} to specify $\Mbh$.  We
still need to specify $i$ and $a_*$.  For all but the highest masses,
the inclination has the strongest effect on the derived accretion
rate, primarily due to the $\cos i$ dependence of the projected disk
area.  It is reasonable to assume that $\cos i\sim 0.5-1$ as nearly
edge on systems are likely obscured.  Thus, we adopt $\cos i=0.8$ as a
characteristic value. We examine the implications of inclination
dependence in later sections.

Figure \ref{f:spin_comp} shows the effect that varying $a_*$ has on
relativistic blackbody models.  At fixed $\Mbh$ and $\Mdot$, the peak
of the spectrum increases with $a_*$.  Since $\Mdot$ is fixed, $L_{\rm
  bol}$ increases with $a_*$ due to the increase in efficiency.
However, this increase in $L_{\rm bol}$ mostly manifests itself as
increased emission at high frequencies.  At visible frequencies, the
change with spin is much more modest.  At $10^7 \Msun$ varying $a_*$
has a negligible effect.  For $10^9 \Msun$ there is a more pronounced
effect because the optical photons are typically emitted much closer
to $r_{\rm ms}$ than at lower $\Mbh$, but the variation in predicted
$L_{\rm opt}$ from $a_*=0-0.998$ is only $\sim 50\%$.

\begin{figure}
\includegraphics*[width=\columnwidth]{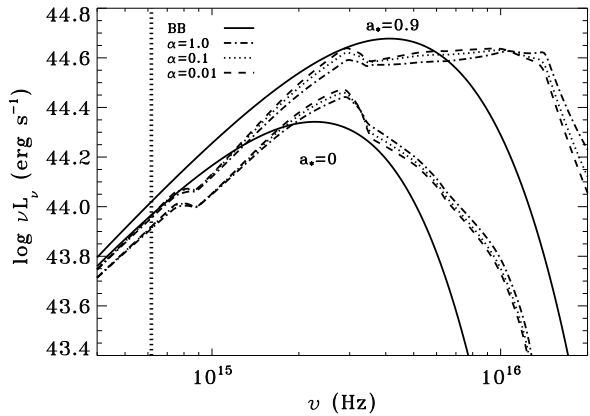}   
\caption{
Comparison of local blackbody AD SEDs with the detailed atmospheric model 
TLUSTY, for different values of $a_*$ and $\alpha$. All models have
$\Mdot =0.1 \Msun \rm \; yr^{-1}$, $\Mbh=10^8 \Msun$, and $\cos i=0.8$. 
Note the increasing atmospheric effects in the hotter parts of the disk, which
are sensitive to the unknown viscosity mechanism (although the effect
of $\alpha$ appears small). However, $L_{\rm opt}$ is rather insensitive to 
the atmospheric structure and radiative transfer effects, as it originates from
colder parts of the AD, which is expected to emit locally close to a blackbody.
Thus, $L_{\rm opt}$ should be mostly set by $M\Mdot$ (see eq. \ref{eq:lnu2}).
\label{f:mod_comp}}
\end{figure}

Figure \ref{f:mod_comp} shows the differences between TLUSTY models
($\alpha_{\rm SS}=0.01-1.0$), and blackbody models computed with the
same $\Mdot$ and $\Mbh$ for $a_*=0$ and 0.9.  The TLUSTY calculations
tend to produce spectra which are nearly blackbody in the optical, but
there is still a contribution to $L_{\rm opt}$ from the low energy
tails of somewhat hotter annuli.  These tend to be modified
blackbodies (but with imprints of the Balmer edge) due to electron
scattering, which shifts power to higher frequencies, leaving less
flux at optical frequencies relative to the blackbody prediction.  At
UV frequencies, this effect is more pronounced for higher $a_*$, but
the decrement is relatively independent of spin in the optical
emission.

For this $\Mdot$ and $M$, the choice of $\alpha_{\rm SS}$ has very
little impact at any frequency, although the differences are larger at
higher frequencies.  This is due to the fact that the hottest, inner
annuli tend to be the lowest surface density in a \cite{ss73} disk
model. As discussed in \cite{dd08}, the spectra are generally
insensitive to surface density as long as it is sufficiently large so
that the disk is very optically thick. For higher Eddington ratio
models, the surface density is low enough that $\alpha_{\rm SS}$ can
have a significant impact on the UV spectrum \citep{hub00}. However,
for the Eddington ranges of interest, the optically emitting annuli in
AGN disks have large enough surface density that $\alpha_{\rm SS}$ has
very little impact on the optical spectrum.

\begin{figure}
\includegraphics*[width=\columnwidth]{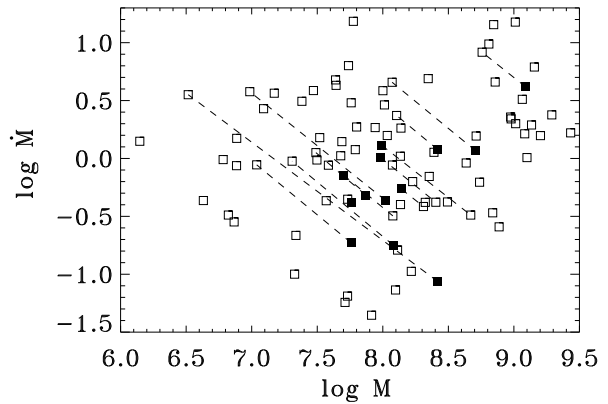}   
\caption{
The derived $\Mdot$ (in $\Msun \rm \; yr^{-1}$) as a function of $M$
for the 80 PG QSO in our sample.  The $a_*=0.9$ and $\cos
i=0.8$ TLUSTY model is used to derive $\Mdot$. The open and filled
squares are derived using $M_{\rm BLR}$ and $M_{\sigma}$,
respectively.  The typical discrepancy in $M$, and in the implied $\Mdot$,
is a factor of $2-3$, but can reach a factor of 10. Interestingly, the derived
$\Mdot$ is only weakly correlated with $M$, although the AGN luminosity 
rises more steeply with $M$ in our sample (see below).
\label{f:mdot}}
\end{figure}

\subsection{Accretion Rates}

With $a_*$, $i$, $M$, and $\alpha_{SS}$ specified, estimating $\Mdot$
is straightforward.  We first calculate $L_{\rm opt}$ for models with
the above parameters and different $\Mdot$.  We then linearly
interpolate to find the $\Mdot$ for which the model optical flux would
match the observed value.  The resulting rates for our sample of PG
quasars are shown in Figure \ref{f:mdot} and reported
in Table~1 and Table~2.  A TLUSTY model with
$a_*=0.9$ and $\cos i=0.8$ was used to obtain these estimates.

The $\Mdot$ estimates are sensitive to the assumed
$\Mbh$ and the differences between $M_{\rm BLR}$ and $M_{\sigma}$ can be
significant.  Consistent with equation (\ref{eq:mdot}), a larger
$\Mbh$, yields a smaller $\Mdot$ and vice versa.  The factors of 2-3
discrepancies in $\Mbh$ yield comparable uncertainties in $\Mdot$.

\begin{figure}[h]
\includegraphics*[width=\columnwidth]{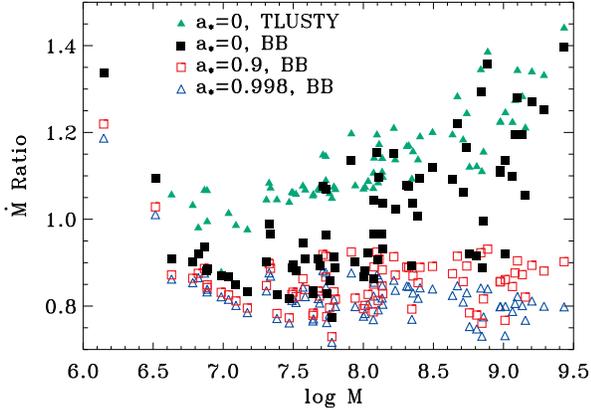}   
\caption{
The effect of different assumptions about $a_*$ and the local
disk spectrum on the derived $\Mdot$.  For each source in our sample, we compute $\Mdot$ as
described in \S\ref{model} using five different disk models.  All
values of $\Mdot$ are plotted relative to the $a_*=0.9$, the TLUSTY
based model results shown in Figure \ref{f:mdot}. The uncertainty in
$\Mdot$ is generally well below 40\%, and is negligible compared to the
errors resulting from the uncertainty in $M$.
\label{f:mdot_ratio}}
\end{figure}

Figure \ref{f:mdot_ratio} shows how $\Mdot$ depends on $a_*$ and the
method used to compute the surface emission.  For each QSO in our
sample, we compute five values of $\Mdot$ using different models for
the SEDs.  The base model corresponds to the $a_*=0.9$ TLUSTY model
used to derive the accretion rates plotted in Figure \ref{f:mdot}.
The other four $\Mdot$ estimates are used to compute ratios with the
base model $\Mdot$ in the denominator.  The TLUSTY model with $a_*=0$,
generally provide a higher $\Mdot$.  This is a 10\% effect at low
$\Mbh$, but can reach 40\% at the highest $\Mbh$.  Blackbody models
with $a_*=0$ also tend to give similarly higher $\Mdot$ for larger
$\Mbh$, but can be 20\% lower at low $\Mbh$.  Blackbody models with
higher spin, $a_*=0.9$ and $a_*=0.998$, give about 10-20\% lower
$\Mdot$ for almost all $\Mbh$.

The objects with the two lowest masses correspond to very high
Eddington ratios, which creates difficulties for the TLUSTY models.
This is due to the fact that the annuli computed directly with TLUSTY
do not cover the parameter range needed for the spectral models, and
extrapolation (rather than interpolation) is used to construct the
spectra.  As a result, the $\Mdot$ obtained with TLUSTY is probably an
underestimate, resulting in the higher ratio of the blackbody models
relative to this model that can seen in Figure \ref{f:mdot_ratio}.

The differences in $\Mdot$ derived using different $a_*$ are
consistent with Figure \ref{f:spin_comp} in that higher $a_*$ yield
larger $L_{\rm opt}$ when $\Mbh$ is large, but comparable $L_{\rm
  opt}$ when $\Mbh$ is small.  A larger model $L_{\rm opt}$ means that
the observed $L_{\rm opt}$ can be matched with a lower $\Mdot$.  The
difference between TLUSTY and blackbody models are similarly
consistent with Figure \ref{f:mod_comp} in that TLUSTY models
generally give lower $L_{\rm opt}$, and thus require a higher $\Mdot$
to match observations. The uncertainty in $a_*$, coupled
with differences in the spectral models, corresponds to an overall
uncertainty in $\Mdot$ of $\sim 20\%$ at $\Mbh=3 \times 10^{6} \Msun$
and $\sim 40\%$ at $\Mbh=3 \times 10^{9} \Msun$.  We will find that
this is significantly lower than the uncertainty associated with the
$\Mbh$ estimates.

We are now in a position to evaluate how well our simple analytic
relation (\S\ref{simple}) approximates our more sophisticated fitting
method (\S\ref{model}).  Treating the logarithms of either ($L_{\rm opt}$,
$\Mbh$) or ($L_{\rm opt}$, $v_{3000}$) as sets of independent
variables, we perform a linear least-squares fit to our $\Mdot$
estimates.  Our best fit relations are
\begin{equation}
\Mdot = 3.5 \;\Msun \; {\rm yr^{-1}} \; M_8^{-0.89} L_{\rm opt, 45}^{1.5},
\label{eq:mdotfit1}
\end{equation}
and
\begin{equation}
\Mdot = 2.5 \;\Msun \; {\rm yr^{-1}} \; v_{3000}^{-1.78} L_{\rm opt, 45}^{0.87},
\label{eq:mdotfit2}
\end{equation}
Equation \ref{eq:mdotfit1} is useful when $M_\sigma$ is available, and 
equation \ref{eq:mdotfit2} when only $M_{\rm BLR}$ is available.

\begin{figure}[h]
\includegraphics*[width=\columnwidth]{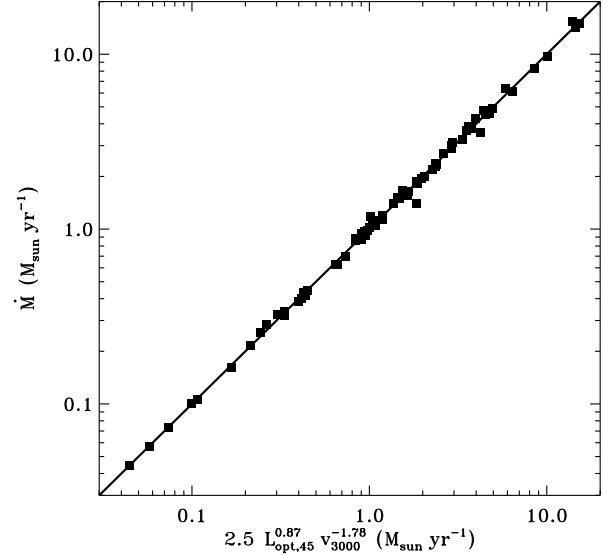}   
\caption{
The model fit derived $\Mdot$ plotted versus the best fit analytic
relations (eq.\ref{eq:mdotfit2}) using
$L_{\rm opt}$ and $v_{3000}$ as independent variables. The RMS deviation
of $\Mdot$ from the analytic fit is $\Delta \log \Mdot=0.024$.
This analytic relation provides a simple and accurate estimate for $\Mdot$,
which does not require model fitting. 
\label{f:mdotfit}}
\end{figure}

As shown in figure \ref{f:mdotfit}, equation \ref{eq:mdotfit2} 
provides a very good fit to
our model-based $\Mbh$ estimates, with an RMS error of $\Delta
\log \Mdot=0.024$.  Such precise agreement suggests that
equation (\ref{eq:mdotfit1}) or (\ref{eq:mdotfit2}) could be used in
place of a detailed model fitting method for future work.  A
comparison of equations (\ref{eq:mdot}) and (\ref{eq:mdotfit1})
demonstrates reasonable agreement between the simple analytic model
and the best-fit relations, although the dependence on $\Mbh$ is
somewhat flatter than naively expected.  This slight discrepancy is
due to the effects of the inner boundary, which becomes more
pronounced as $\Mbh$ increases and the radius of optical emission gets
closer to the inner boundary.

\section{Estimating the Radiative Efficiency}
\label{efficiency}

\subsection{Bolometric Luminosity Estimates}
\label{lbol}

Estimation of the radiative efficiency of an accretion flow clearly
requires a reliable measurement of the bolometric luminosity which is
radiated.  Since broadband constraints on the SED are a priority, we
focus on a sample of 80 relatively well observed PG quasars
\citep{bg92,bl05}, for which optical, UV, and X-ray data are
available.  However, the SED in the extreme UV, where the AD
spectrum is expected to peak, remain unknown.

\begin{figure*}[t]
\includegraphics*[width=\textwidth]{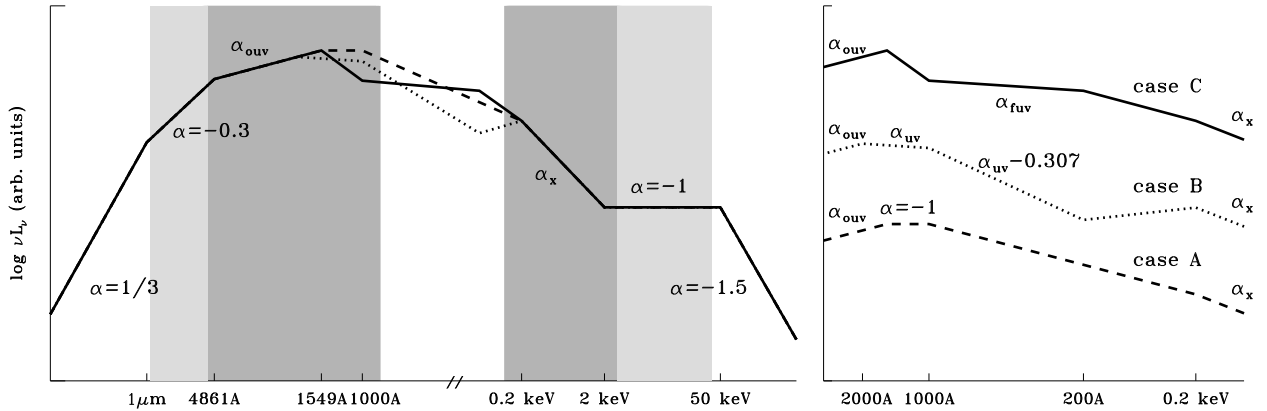}
\caption{
Schematic piecewise power law SED used to estimate $L_{\rm bol}$ in
our sample.  In the left panel we plot representative case A (dashed),
case B (dotted), and case C (solid) model SEDs as described in the
text.  The dark shaded areas denote frequency ranges where the SED is
computed using observations for individual models, while
the light shaded areas show frequencies where we have assumed
characteristic slopes motivated by QSO observations at these
frequencies.  The unshaded regions are interpolations or
extrapolations.  Note that the optical to UV frequency range is
expanded relative to X-ray for clarity. The right panel is a
`zoomed-in' plot of the UV to soft X-ray region of the left panel to
highlight differences between the three cases of model SEDs.  The three
different possible cases for the SED interpolations imply a typical
(RMS) uncertainty of 0.24 in $\log L_{\rm bol}$.
\label{f:sed}}
\end{figure*}

We compile data on the SED from several sources. The redshift, fluxes
at 1549~\AA~ and 4861~\AA, the power-law slopes between these
wavelengths $\alpha_{\rm ouv}$, and the H$\beta$ FWHM are all taken
from \cite{bl05}.  We also use a UV spectral slope $\alpha_{\rm uv}$
computed between 2000~\AA~ and 1400~\AA~ by A. Baskin (private
communication).  Since these values are not published elsewhere, we
report them in the fourth column of Table~1. If available, the far UV
slopes $\alpha_{\rm fuv}$ and 1000~\AA~ fluxes are taken from analysis
of FUSE data presented in \citet{sco04} and \citet{sha05}\footnote{For
  sources listed in both \citet{sco04} and \citet{sha05}, we use the
  \cite{sha05} values.}.  The soft X-ray slope $\alpha_{\rm x}$ (0.2-2
keV) is computed from the H$\beta$ FWHM using the relation in
\citet{blw00}.  The flux at 1 keV is taken from \citet{lb08}, who
tabulated the data of \citet{blw00} and \citet{lb02}.  All power-law
slopes described in this text are $\alpha_\nu$ (i.e. $F_\nu \propto
\nu^\alpha$).  Fluxes are converted to luminosities using the
redshifts listed in \citet{bl05} and assuming a $\Lambda$CDM
cosmology with $H_0=70 \;{\rm km \; s^{-1} \; Mpc^{-1}}$,
$\Omega_m=0.3$, and $\Omega_{\Lambda}=0.7$.

With these data we compute $L_{\rm bol}$ using a piece-wise power law
representation for the SED.  To clarify the discussion that follows,
we plot a characteristic example SED in Figure \ref{f:sed}.  It is
relatively straightforward to estimate the continuum for wavelengths
longer than 1549~\AA~ and shorter than 62~\AA~ (0.2 keV).  First, we
exclude the observed infrared bump shortward of 1 $\mu$m.  This
emission is thought to be reradiated by dust and we are only
interested in the direct emission from the AD.  Therefore, we assume
power-laws with $\alpha=1/3$ below 1 $\mu$m and $\alpha=-0.3$ between
1 $\mu$m and 4861~\AA.  Between 4861~\AA~ and 1549~\AA, we assume a
power law slope equal to the measured $\alpha_{\rm ouv}$.  The
spectrum between 0.2 and 2 keV is assumed to have a power-law slope
equal to $\alpha_{\rm x}$ and is normalized to match the 1 keV flux.
From 2 keV to 50 keV, we assume $\alpha=-1$, and above 50 keV we
assume $\alpha= -1.5$.

We consider three different prescriptions for the unobserved extreme UV spectrum.
In case A, we assume a power-law with $\alpha=-1$ between 1549~\AA~
and 1000~\AA, and another power-law is fit through 1000~\AA~ and 0.2
keV.  In case B, we use $\alpha_{\rm ouv}$ for the power law slope
between 4861~\AA~ and 2000~\AA, but use the measured $\alpha_{\rm uv}$
from 2000~\AA~ to 1000~\AA.  Between 1000~\AA~ and 200~\AA, we assume
$\alpha=\alpha_{\rm uv}-0.307$, and a power-law is fit between
200~\AA~ and 0.2 keV.  Finally, for those sources with FUSE data, we
also consider a case C. We fit a power law between the 1549~\AA~ and
1000~\AA~ fluxes.  From 1000~\AA~ to 200~\AA, we use $\alpha_{\rm
  fuv}$, and a power-law is again fit between 200~\AA~ and 0.2 keV.
Note that $\Delta \alpha=-0.307$ in case B corresponds to the mean
difference between $\alpha_{\rm uv}$ and $\alpha_{\rm fuv}$ in those
sources where $\alpha_{\rm fuv}$ is available from FUSE data.

\begin{figure*}
\includegraphics*[width=0.75\textwidth]{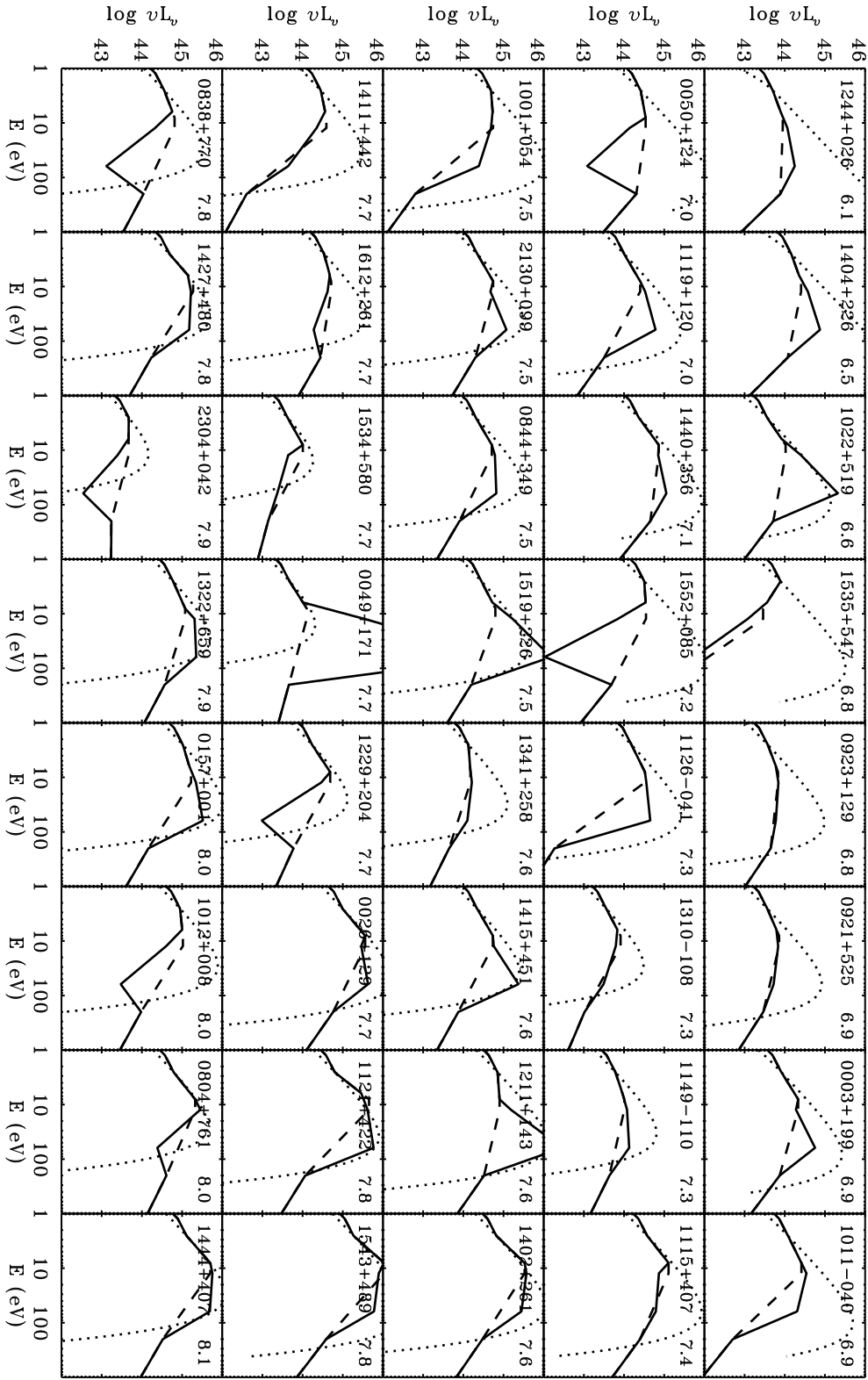}
\caption{
The SED derived for each object. The solid curve is the SED
used to compute $L_{\rm bol}$.  It corresponds to case C for sources
with FUSE data and case B otherwise.  The dashed curve shows the case
A SED for comparison.  These SEDs are described in detail in
\S\ref{lbol}. The dotted curve is the $a_*=0.9$ local
blackbody AD model with $\Mbh = M_{\rm BLR}$ that matches the SEDs at
4861 \AA. The objects are ordered by ascending order of $M$. Note that
the model fit is hotter than the observed SED for low $M$ objects,
and colder for the high $M$ objects. Thus, the observed AGN SED in our
sample is inconsistent with a fixed $a_*$ AD model 
\label{f:seds}}
\end{figure*}

The resulting SEDs for case A, and for B or C are plotted for each
source in Figure \ref{f:seds}. The plots are arranged in order of
increasing $\Mbh$.  We also show an $a_*=0.9$ relativistic blackbody
model with the known $\Mbh$, and a value of $\Mdot$ which has been fit
to match the luminosity at 4861~\AA~ (see \S\ref{model}).  In most of
the quasars, the SED peaks in the observable far UV (rather than the
unobservable extreme UV).  The X-rays generally contribute only a
modest fraction of the overall luminosity.

For the majority of sources there is reasonable agreement between the
various SED models, but there are several sources with rather
discrepant results.  For cases B and C, there are a number of sources
where $\alpha_{\rm uv}$ or $\alpha_{\rm fuv}$ seem to be rising or
falling much more steeply than appears likely.  This is presumably due
to some combination of slope and flux measurement errors, dust
reddening, and variability (the X-ray and UV fluxes are not
contemporaneous).  In sources with FUSE observations there is
sometimes a mismatch between the luminosity estimates at 1000~\AA~ and
the \citet{bl05} data at 1549~\AA, presumably due to instrumental
uncertainties and variability.

\begin{figure*}
\includegraphics*[width=0.75\textwidth]{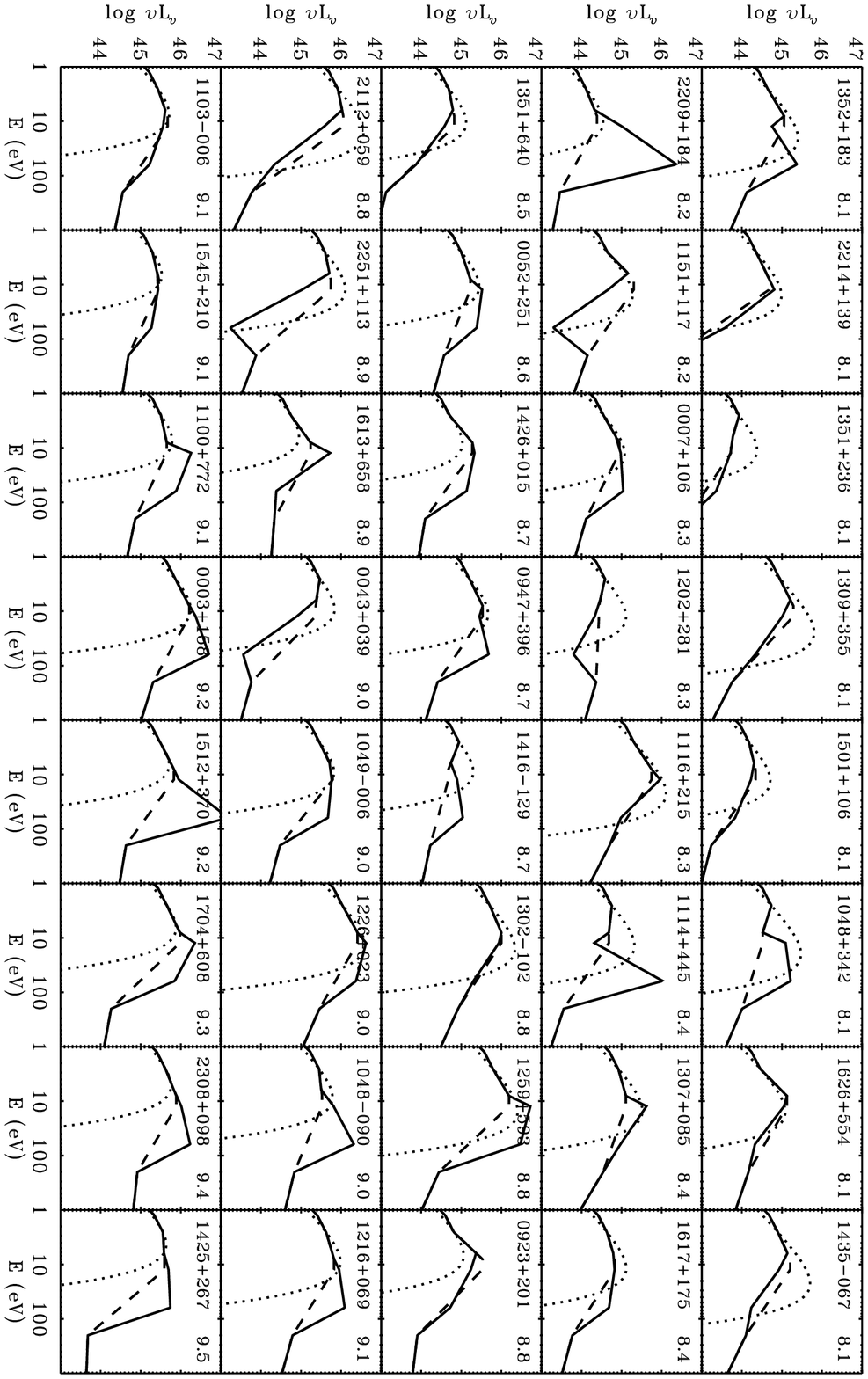}
\begin{flushleft}
Fig.~\ref{f:seds}. (continued)
\end{flushleft}
\end{figure*}

The case A SED, constructed to form a smooth transition between the UV
and soft X-rays, often seems to be the ``most reasonable''.  However,
it is important to avoid enforcing a universal SED preconception, and
so we adopt the approach of using as much UV data as possible, even
when it gives apparently unlikely FUV extrapolations.  Therefore, in
sources where FUSE data is available, we use the case C SED to
estimate $L_{\rm bol}$.  In sources where it is not available, we use
case B.  We report the resulting $L_{\rm bol}$ estimates in the fifth
column of Table~1.  We find that this choice increases the scatter in
our $\eta$ estimates over what we would have obtained using case A,
but does not significantly change the overall trend with $\Mbh$.  We
make one exception for PG 0049+171, which has $\alpha_{\rm fuv}=4.1$.
Such an unphysically steep slope leads to an extreme overestimate of
$L_{\rm bol}$, so we neglect it in all further analysis.

There are several potential sources of uncertainty in the $L_{\rm
  bol}$ estimate derived from these SEDs, but in the vast majority of
sources the dominant uncertainty is the far-UV extrapolation.  To
estimate this uncertainty, we assume $\Delta \alpha=\pm 1$ for far-UV
extrapolation $\alpha_{\rm fuv}$ (which is equal to $\alpha_{\rm
  uv}-0.307$ for case B) and compute the resulting range of $L_{\rm
  bol}$. This is considerably greater than the typical measurement
errors in the sources where $\alpha_{\rm fuv}$ is measured directly
with FUSE.  We adopt a larger uncertainty to conservatively account
for any systematic errors, possible effects of dust reddening, and
potential complexity in the unobserved part of the SED.  Since this
generally yields a large uncertainty that dominates other sources of
error (e.g optical or X-ray variability) we assume the contribution
from other sources is negligible.

Note that in Figure \ref{f:seds}, the AD model, which has a fixed
$\eta=0.16$ ($a_*=0.9$), systematically overpredicts the observed FUV
SED in the lowest $\Mbh$ objects and produces a higher overall $L_{\rm
  bol}$.  In contrast, the AD model systematically underpredicts the
observed FUV SED for the highest $\Mbh$ objects, and produces a lower
total $L_{\rm bol}$.  This systematic trend suggests that a single
value of $\eta$ will not be consistent with all objects, but instead
implies that $\eta$ needs to increase from low to high $\Mbh$, as we
discuss below.

\subsection{Radiative Efficiencies}
\label{eta}

Now that we have estimates for $L_{\rm bol}$ and $\Mdot$ we can
estimate $\eta$ using
\begin{equation}
\eta=\frac{L_{\rm bol}}{\Mdot c^2 \cos i}.
\label{eq:lbol}
\end{equation}
The factor of $\cos i$ accounts for the inclination dependence of the
observed $L_{\rm bol}$, if it originates in a thin AD.\footnote{For a
  flat Newtonian AD there is an additional factor of 2 in the
  numerator. However, relativistic beaming and electron scattering
  induced limb darkening modify the angular distribution at higher
  frequencies, invalidating the Newtonian approximation. We therefore
  neglect the factor 2 for simplicity.}  Here we assume $\cos i=0.8$
as noted above (\S 2.3).

\begin{figure}[h]
\includegraphics*[width=\columnwidth]{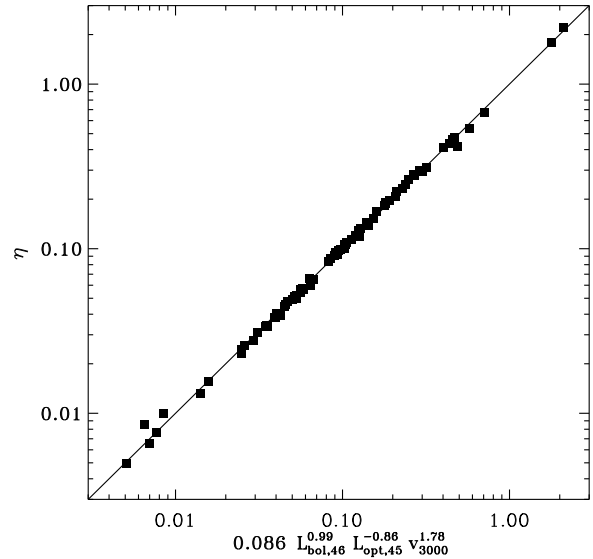}
\caption{
The derived $\eta$ plotted versus the best fit analytic relations (eq.
\ref{eq:etafit2}) using $L_{\rm opt}$, $L_{\rm bol}$, and $v_{3000}$
as independent variables.  The RMS deviation of $\eta$ from the
analytic fit is $\Delta \log \eta=0.024$, i.e.  an accuracy of 5\%.
\label{f:eta_fit}}
\end{figure}

A linear least squares fit for $\log \eta$ as a function of 
the logarithms of $\Mbh$, $L_{\rm opt}$, assuming a constant error
for each $\eta$ estimate, yields best fits of
\begin{equation}
\eta=0.063 L_{\rm bol,46}^{0.99} L_{\rm opt,45}^{-1.5} M_8^{0.89},
\label{eq:etafit1}
\end{equation}
or
\begin{equation}
\eta=0.086 L_{\rm bol,46}^{0.99} L_{\rm opt,45}^{-0.86} v_{3000}^{1.78},
\label{eq:etafit2}
\end{equation}
with rms deviations of $\Delta \log \eta=0.024$ from the true values.
A comparison of these best-fit values with our TLUSTY based model
estimates is shown in Figure \ref{f:eta_fit}.
We would have arrived at nearly identical relations if we had simply
inserted equations (\ref{eq:mdotfit1}) and (\ref{eq:mdotfit2}) into
equation (\ref{eq:lbol}).  The best fitting relation when $L_{\rm
  bol}$ is not available is
\begin{equation}
\eta=0.08 L_{\rm opt,45}^{-0.42} M_8^{0.83} ,
\end{equation}
with an rms deviation of $\Delta \log \eta=0.26$. This 
is still interesting, but considerably poorer than the three
parameter fit when $L_{\rm bol}$ is available.

To understand these best-fit relations we look at the expected
relation for the simple Newtonian AD case.
Inserting equation (\ref{eq:mdot}) into (\ref{eq:lbol}) yields
\begin{equation}
\eta =0.068 L_{\rm bol,46} L_{\rm opt,45}^{-3/2} M_8 (\cos i)^{1/2},
\label{eq:etaa}
\end{equation}
or using equation (\ref{eq:mass})
\begin{equation}
\eta =0.11 L_{\rm bol,46} L_{\rm opt,45}^{-0.81} v_{3000}^2 (\cos i)^{1/2}.
\label{eq:etab}
\end{equation}
The assumption that $\cos i \sim 0.5-1$, combined with the $(\cos
i)^{1/2}$ dependence suggest that inclination uncertainties are not
likely to be a significant source of uncertainty in $\eta$.  With
the inclination dependence removed, these results are in approximate
agreement with our best fit relations, confirming that simple
Newtonian blackbody model captures most of the relevant physical 
effects.

\begin{figure}[h]
\includegraphics*[width=\columnwidth]{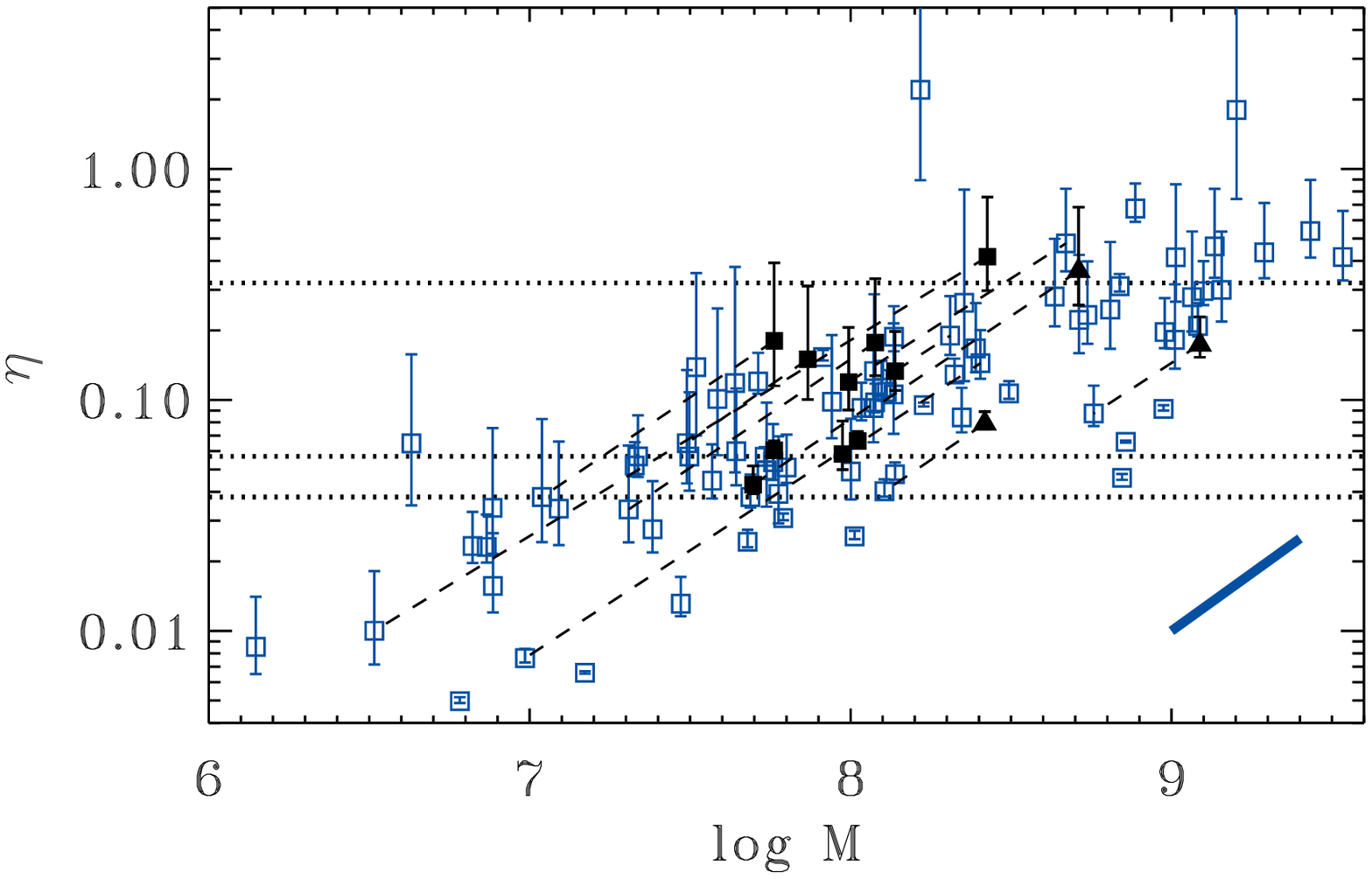}
\caption{
A plot of $\eta$ versus $\Mbh$ for the 80 PG quasars in our sample.
The symbols correspond to estimates made with $M_{\rm BLR}$ (blue,
open squares) and $M_{\rm \sigma}$ estimates obtained by \citet{das07}
(filled squares) and \citet{ws08} (filled triangles).  The dashed
solid lines connect the two sets of $\Mbh -\eta$ estimates for the 13
sources for which both $M_{\rm BLR}$ and $M_{\rm \sigma}$ are
available.  The $\eta$ error bars account for uncertainties in $L_{\rm
  bol}$ only.  The horizontal dotted lines correspond to 
$\eta= 0.038, 0.057$ and 0.31, the theoretical efficiencies from
\citet{nt73} for $a_*=-1$, 0, and 0.998, respectively. The thick solid
line in the lower right hand corner of the plot shows the displacement
$\eta - \Mbh$ plane that would occur for a 0.4 dex shift in $\Mbh$.
Two objects, PG 2209+184 and PG 1512+307, have $\eta > 1$, although
their errors are consistent with $\eta \sim 1$. Note that most objects 
fall within the theoretically expected range for $\eta$. The outlying
objects are within the expected uncertainty in $M$. There is a clear 
rise in $\eta$ with $M$, with a best fit relation $\eta=0.089 M^{0.52}$. 
\label{f:eta}}
\end{figure}

In Figure \ref{f:eta} we plot $\eta$ as a function of $\Mbh$ using
$M_{\rm BLR}$ and $M_{\sigma}$ (\citealt{das07}; \citealt{ws08}).  We
report the BLR and $\sigma$ based estimates of $\eta$ in the last
columns of Table~1 and Table~2, respectively.  For the 13 objects with
both $M_{\sigma}$ and $M_{\rm BLR}$ estimates, $\eta$ is plotted for
both cases and the symbols are connected with dashed lines.  The
$\eta$ error bars are computed from the uncertainties in $L_{\rm
  bol}$. The uncertainty associated with the value of $\Mbh$ is likely
larger for most sources, but is not included.  Since the uncertainty
in $\Mbh$ affects both the abscissa and ordinate it will shift points
diagonally in the plot (as seen for the dashed lines).  The thick
solid line in the lower right hand corner of the plot shows the
displacement in the $\eta - \Mbh$ plane that would occur for a 0.4 dex
shift in $\Mbh$.  The impact of this uncertainty is discussed in
detail in \S\ref{mbh}. 

The mean values of $\eta$ are $\log \eta=-1.05 \pm 0.52$ and $-0.91\pm
0.30$ for the estimates made with $M_{\rm BLR}$ and $M_{\sigma}$.
These values are consistent with the values derived from the Soltan
argument (see \S 1) which is based on completely independent
arguments.  In addition, we find a clear correlation of $\eta$ and
$\Mbh$. We find a Spearman rank correlation coefficient $r=0.85$, with
a significance (probability) $P_r=1\times 10^{-19}$ and a best fit
relation
\begin{equation}
\eta = 0.089 M_8^{0.52}
\label{eq:eta_corr}
\end{equation}
The majority of the sources are consistent with $0.057 < \eta <
0.321$, the \citet{nt73} efficiencies for $0< a_* <0.998$.
At high $\Mbh$, several sources have $\eta > 0.321$, but are generally
also consistent with $\eta\le 0.321$.  There are two sources with
$\eta>1$ (PG 2209+184 at $\log \Mbh=8.2$ and PG 1512+307 at $\log
\Mbh=9.2$) and Figure \ref{f:seds} shows that both have two of the
most steeply rising FUV slopes in the sample.  These FUV slopes are
difficult to reconcile with the X-ray flux level and are likely
erroneous, leading to overestimation of $L_{\rm bol}$ and, therefore,
overestimation of $\eta$.  In contrast, At low $\Mbh$, there are
numerous sources with relatively small error bars, based on the
$L_{\rm bol}$ estimates, for which $\eta$ is significantly less than
0.057. Our result is consistent with the result of Collin et
al. (2002), who assumed a constant $\eta$ model, and derived an
increasing $L_{\rm bol}/L_{\rm Edd}$ with decreasing $M$, with $L_{\rm
  bol}/L_{\rm Edd}$ reaching 1-100 for an assumed $\eta=0.32$. Assuming
$L_{\rm bol}/L_{\rm Edd}$ is at most only
slightly greater than unity, as inferred for our sample, this implies
$\eta\sim 0.01$ for the low $M$ objects in their sample, broadly
consistent with our result.

\begin{figure}
\includegraphics*[width=\columnwidth]{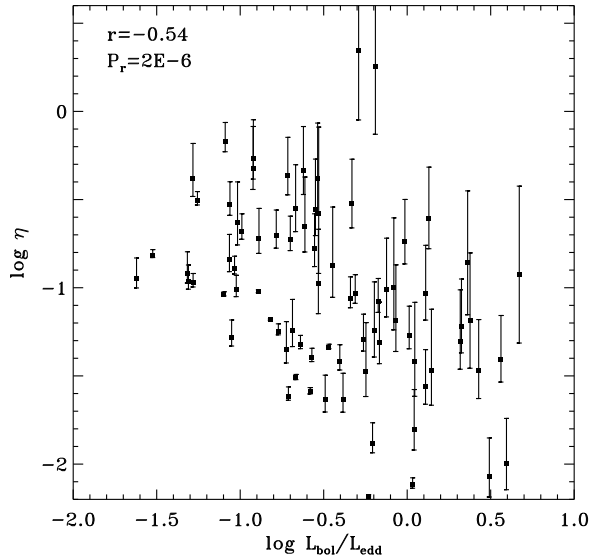}
\caption{
Correlation of $\eta$ with Eddington ratio. The Spearman rank
correlation coefficient $r$, and corresponding significance $P_r$
confirm that a modest correlation exists. However, this correlation is
likely induced by the stronger correlation of $\eta$ and $M$
(Fig.\ref{f:eta}), and the correlation of the Eddington ratio with $M$
(see below).
\label{f:eta_ledd}}
\end{figure}

\begin{figure}
\includegraphics*[width=\columnwidth]{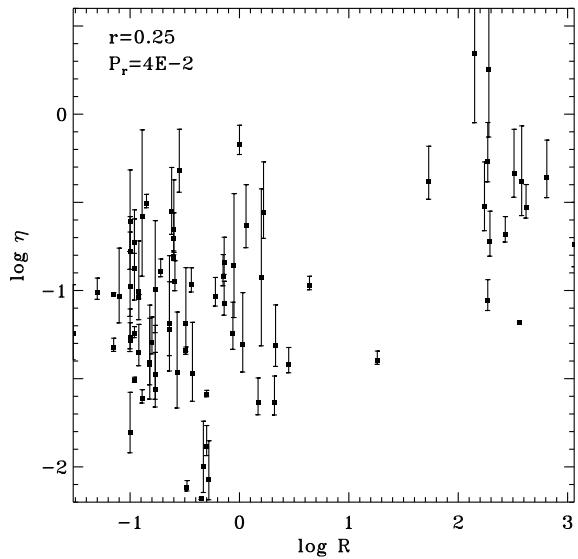}
\caption{
Correlation of $\eta$ with radio loudness, $R$. The Spearman rank
correlation coefficient $r$, and corresponding significance $P_r$
indicate that only a weak correlation exists. This correlation is
likely induced by the correlation of $\eta$ and $M$, and the known
correlation of radio loudness and $M$ in the PG quasars.
\label{f:eta_r}}
\end{figure}

In Figures \ref{f:eta_ledd} and \ref{f:eta_r}, we plot the variation of the $M_{\rm BLR}$
based $\eta$ with $L_{\rm bol}/L_{\rm Edd}$ and the radio
loudness parameter $R\equiv f_{\rm 6cm}/f_{4400}$ from \cite{lb08}.
In each panel we provide $r$ and
$P_r$ for each distribution. Figures \ref{f:eta_ledd} shows a modest
($r=0.54$) correlation of $\eta$ with $L_{\rm bol}/L_{\rm Edd}$.
In Figure \ref{f:eta_r} there is a tendency for
radio loud quasars to have higher average $\eta$ than radio quiet
sources, although the distributions overlap and the resulting
correlation is weak ($r=0.25$). Given that Figure \ref{f:eta} shows a
correlation of $\eta$ with $\Mbh$, such a tendency is not unexpected
since there is a well-known correlation of $R$ with $\Mbh$
\citep[e.g.][]{lao00}.  This is in qualitative agreement with 
the results of \citet{bz03} of a higher $\eta$ for the radio loud
quasars.

\subsection{$L_{\rm opt}/L_{\rm bol}$ Ratio and its Implications}
\label{lum_rat}

\begin{figure}
\includegraphics*[width=\columnwidth]{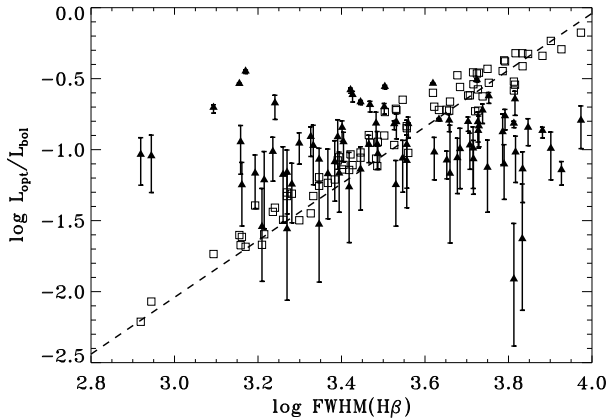}
\caption{
Ratio of $L_{\rm opt}/L_{\rm bol}$ as a function of the FWHM of
H$\beta$ for the objects in our sample.  The errors are computed
assuming an uncertainty of $\Delta \alpha=\pm 1$ for $\alpha_{\rm
  fuv}$ for all objects.  The dashed line is a simple analytic
estimate (eq. \ref{eq:rat}) for constant $\eta$.  The open squares are
also constant $\eta$ estimates, but instead based directly on our
$\Mdot$ estimates, as described in \S\ref{lum_rat}. The ratio of
$L_{\rm opt}/L_{\rm bol}$ shows a relatively small spread, and no
correlation with the FWHM of H$\beta$, and demonstrates again that
constant $\eta$ thin AD models are excluded.
\label{f:t0}}
\end{figure}

In Figure \ref{f:t0} we plot the ratios $L_{\rm opt}/L_{\rm bol}$
versus the FWHM of the H$\beta$ line.  The observed ratio of $L_{\rm
  opt}/L_{\rm bol}$ shows almost no trend with the FWHM. The
distribution of $\log L_{\rm opt}/L_{\rm bol}$ has a mean of -0.96
with a standard deviation of 0.27.  The errors are generally larger
for objects with low $L_{\rm opt}/L_{\rm bol}$.  In these objects, the
far UV tends to account for a larger fraction of $L_{\rm bol}$, so the
uncertainty associated with the extrapolation into the extreme UV and
soft X-rays tends to be more important, giving a larger overall
uncertainty in $L_{\rm bol}$.  When we instead use the more uniform
case A SED (not plotted), we find considerably lower scatter, but the
absence of a trend with FWHM remains.

The strong correlation of $\eta$ and $\Mbh$ is consistent with the
lack of a correlation between the ratio $L_{\rm opt}/L_{\rm bol}$ and
FWHM in Figure \ref{f:t0}.  Solving  equation (\ref{eq:etab}) for
$L_{\rm bol}$, we obtain
\begin{equation}
\frac{L_{\rm opt}}{L_{\rm bol}} = 0.098 \eta_{0.1}^{-1} L_{\rm opt,45}^{0.19} v_{3000}^2,
\label{eq:rat}
\end{equation}
where $\eta_{0.1}=\eta/0.1$ and we have assumed $\cos i =0.8$.
Therefore, if $\eta$ were constant, this ratio should vary roughly as
the square of the FWHM with only a very weak dependence on $L_{\rm
  opt}$.  We plot this relation as a dashed curve in Figure
\ref{f:t0}.  For simplicity, we ignore the $L_{\rm opt}$ dependence
and set $L_{\rm opt,45}=1$ in equation (\ref{eq:rat}).  Alternatively,
we can use our $\Mdot$ estimates derived from model fitting to specify
$L_{\rm bol}$ via equation (\ref{eq:lbol}) assuming constant
$\eta=0.089$ and $\cos i =0.8$.  These estimates are plotted as open
squares in Figure \ref{f:t0} and follows the dashed curve with little
scatter.  This shows that the evolution of $L_{\rm opt}/L_{\rm bol}$
derived from the TLUSTY models is largely captured by our simple
analytic estimate.  In either case, the expected evolution of $L_{\rm
  opt}/L_{\rm bol}$ for constant $\eta$ is not observed.

\begin{figure}
\includegraphics*[width=\columnwidth]{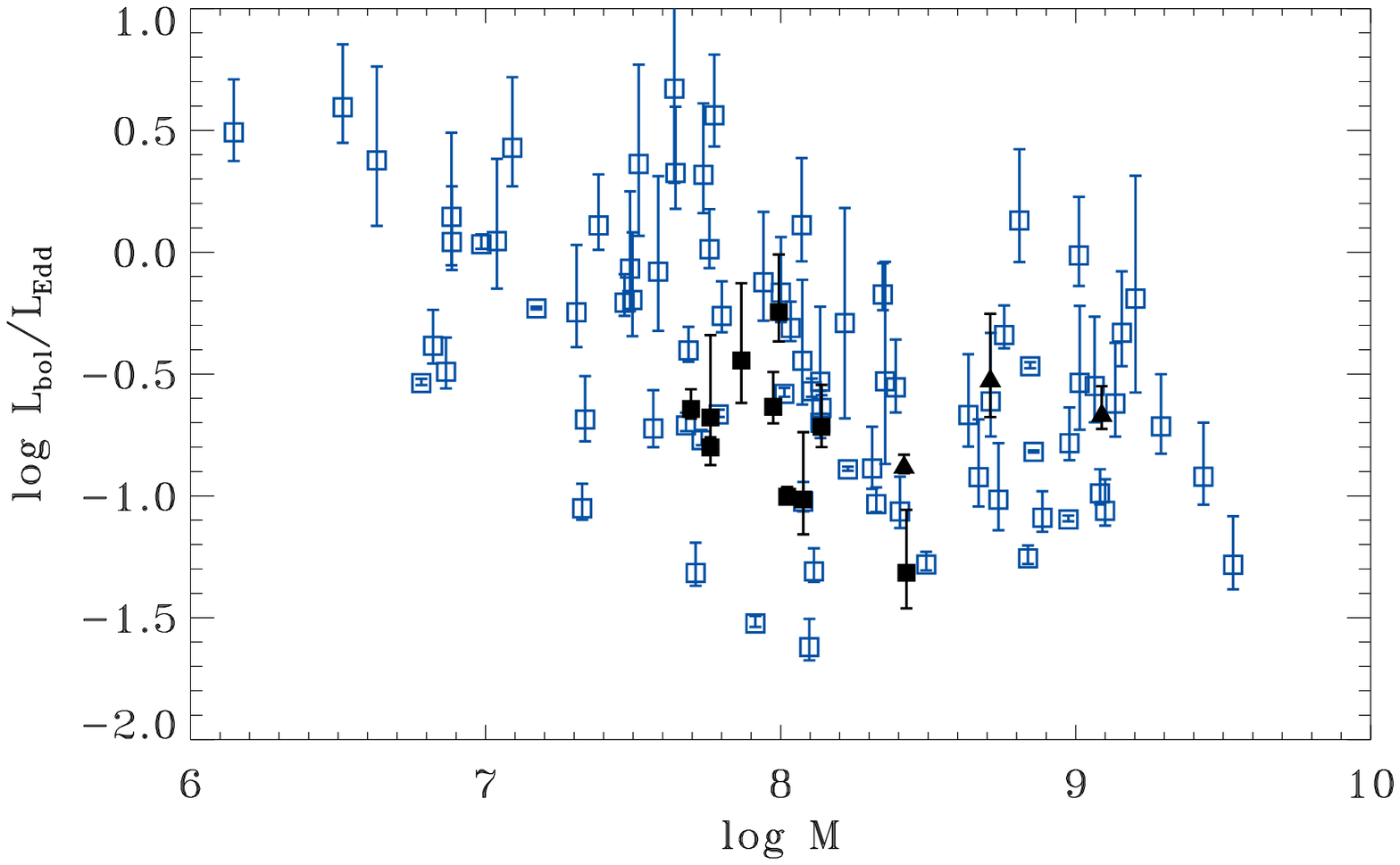}
\caption{
Ratio of $L_{\rm bol}/L_{\rm Edd}$ versus $\Mbh$ for all 80 PG quasars
in our sample.  The symbols correspond to estimates made with $M_{\rm
  BLR}$ (blue, open squares) and $M_{\rm \sigma}$ estimates obtained
by \citet{das07} (filled squares) and \citet{ws08} (filled triangles).
The y error bars account for uncertainties in $L_{\rm bol}$ only.  The
deficiency of objects with $L_{\rm bol}/L_{\rm Edd}<0.1$ likely
results from the selection of point-like objects for the PG sample, as
$L_{\rm Edd}$ is proportional to the host bulge luminosity. The rise
of the lower limit to $L_{\rm bol}/L_{\rm Edd}$ in the lowest $M$
objects may be due to $M$ estimate errors, if all low $M$ objects have
a higher $M$. Or, it could also be a selection effect, if low $M$
objects reside in disk galaxies, where the AGN needs to outshine the
disk light as well.
\label{f:ledd}}
\end{figure}

The above relation assumes that $\Mbh$ is simply a function of $L_{\rm
  opt}$ and FWHM.  Since there is a potential for systematic
uncertainties in these $\Mbh$ estimates, it is useful to consider what
relationship between $L_{\rm bol}$ and $\Mbh$ would be needed to
reconcile a nearly constant $L_{\rm opt}/L_{\rm bol}$ with a constant
$\eta$.  From equation (\ref{eq:etaa}), we see that $\Mbh \propto
L_{\rm bol}^{1/2}$ or, equivalently, $L_{\rm bol}/L_{\rm Edd} \propto
\Mbh$ is required to keep $\eta$ constant.  The point like selection
criterion for the PG sample selects against low $L_{\rm bol}/L_{\rm
  Edd}$ objects, where the host galaxy becomes dominant. This likely
induces the relatively high values and small spread in $L_{\rm
  bol}/L_{\rm Edd}$ (Figure \ref{f:ledd}), yielding $L_{\rm bol} \propto \Mbh$.
Thus $\eta\propto \Mbh^{0.5}$, as observed, results from the small
spread in both $L_{\rm bol}/L_{\rm Edd}$ and $L_{\rm opt}/L_{\rm
  bol}$. While the first effect is possibly a selection effect of the
PG sample, the second is almost certainly not.

We note that a relatively uniform ratio of $L_{\rm opt}/L_{\rm bol}$
has been found by previous authors \citep[e.g.][]{elv94,ric06}, and
is, in fact, frequently assumed in the literature to estimate $L_{\rm
  bol}$ when only optical constraints are available.  Using simple
templates with fixed optical-to-UV SED, but accounting for X-ray
variation, \citet{mar04} find weak evolution in the bolometric
correction that corresponds to an increase in $L_{\rm opt}/L_{\rm
  bol}$ with $\Mbh$.  However, this is still a much weaker dependence
than expected for constant $L_{\rm bol}/L_{\rm Edd}$ and $\eta$.
Studies with large QSO samples that assume a constant bolometric
correction and BLR mass estimates generally find $L_{\rm bol}/L_{\rm
  Edd}$ either nearly constant \citep{kol06} or decreasing slightly
\citep{se10} with $\Mbh$.  As with our PG sample, reconciling these
observations with constant $\eta$ not only requires the BLR estimates
to be in error, but $L_{\rm bol}/L_{\rm Edd}$ would have to scale
nearly linearly with $\Mbh$.

\section{Discussion}

In the previous sections we outlined and implemented a prescription
for estimating $\Mdot$ and, ultimately, $\eta$ by assuming ADs are
radiatively efficient.  Our mean $\eta$ is consistent with constraints
derived using the \citet{sol82} argument (e.g.
\citealt{yt02,erz02,mar04,bar05}), and with the theoretically expected 
range of $\eta$ for thin AD. This 
suggests that the radiatively
efficient, thin disk is a viable model for accretion flows in QSOs, at
least at the radii where $L_{\rm opt}$ is generated. However, what
drives the observed strong correlation of $\eta$ with $\Mbh$?  We
consider three basic possibilities: the $\eta-\Mbh$ correlation is
real, possibly due to a correlation of $a_*$ with $\Mbh$; actual
accretion flows differ from the radiative efficient model in an $\Mbh$
dependent manner; or the input parameters ($\Mbh$ or $L_{\rm bol}$)
are not estimated reliably.  We discuss each of these possibilities in
the following subsections.

\subsection{Effects of error in $L_{\rm opt}$ and  $L_{\rm bol}$}
\label{lbolerr}

Our estimates of $L_{\rm bol}$ are subject to two primary
uncertainties: the unknown inclination and our inability to observe in
the the extreme UV.  The unified model of AGNs suggests that our
sample, which are all type 1, are viewed not far from face on
\citep{ant93}.  Evidence suggests that the opening angle increases
with luminosity, but is $\lesssim 60^\circ$ at the higher luminosities
characteristic of our sample \citep{pol08,rey08}.  Therefore we expect
our sample to cover a range of $\cos i \sim 0.5-1$ (or smaller).  To
the extent that the intrinsic emission is isotropic, the geometric
dependence is simply $L_{\rm opt}, L_{\rm bol} \propto \cos i$.  As
noted in \S\ref{eta}, this results in an $(\cos i)^{1/2}$ dependence
for $\eta$, which is only 0.15 dex of scatter for a factor of 2
uncertainty in $\cos i$.  However, this neglects effects due to
atmospheric limb darkening and relativistic beaming, which are
wavelength dependent and, therefore, give different $i$
dependence for $L_{\rm opt}$ and $L_{\rm bol}$.  For the relevant
range of $\Mbh$ and $L_{\rm bol}$, an examination of the model SEDs
suggests a scatter of $\lesssim 0.3$ dex in $\eta$ at fixed $\Mbh$,
for a uniform distribution in $\cos i$ between 0.5 and 1.

Our estimates for $L_{\rm bol}$ and its uncertainty are discussed in
\S\ref{lbol}.  The uncertainties in $L_{\rm bol}$, which determine the
plotted error bars for $\eta$, entirely reflect the uncertainties in
our FUV slope extrapolations.  This uncertainty could account for much
of the scatter in the $\eta-\Mbh$ correlation 
(Figure \ref{f:eta}), but not the overall trend.  A
simple examination of Figure \ref{f:seds} shows why the $\eta-\Mbh$
correlation cannot be easily explained by ``hidden'' emission in the
EUV.  At the high $\Mbh$ end the models predict a rollover in the
spectrum in the observable FUV, which is not seen in most sources.
Thus, any unobserved emission in the EUV, only makes the discrepancy
larger.  At the low $\Mbh$ end, the models continue rising into the
EUV even though there is often a clear flattening or rollover in the
SED in the observed FUV.  Since the X-ray flux is typically below the
FUV, one would require pathologically double peaked SEDs to provide the
``hidden'' emission in these cases.

The fluxes used to construct our SEDs have been corrected for Galactic
dust reddening and neutral absorption, but our analysis does not
account for intrinsic reddening, which could result in a systematic
underestimate of $L_{\rm bol}$ relative to $L_{\rm opt}$ and,
therefore, an underestimate of $\eta$.  An analysis of a large Sloan
Digital Sky Survey QSO sample \citep{hop04} concluded reddening is
modest in most sources, with only a small fraction which are reddened
significantly (i.e. $\lesssim 2$\% with E(B-V) $> 0.1$).  However,
some of the SEDs in Figure \ref{f:seds} give the impression that
reddening might possibly be affecting the FUV and our extrapolation
into the extreme ultraviolet.  Such sources tend to have a smaller
fraction of their emission in the FUV and show up as the lowest $\eta$
sources in their $\Mbh$ bin.  Since the error bars are determined by
the uncertainty of the EUV to soft X-ray contribution to the SED,
which is relatively small for these sources, they also tend to have
relatively small error bars in Figure \ref{f:eta}.  These arguments
suggest that some of the lowest $\eta$ estimates at low $\Mbh$ are
likely underestimates due to reddening.  However, it seems unlikely
that reddening alone can explain the overall correlation of $\eta$
with $\Mbh$, as this would require systematic reddening of all sources
at low $\Mbh$, which is not evident in the SEDs.

Contamination of optical emission from sources other than the AD could
potentially lead to overestimates of $\Mdot$ and underestimates of
$\eta$.  Potential sources of such contamination are host galaxy or
jet emission. However, significant optical jet emission is likely
constrained to the small fraction of core dominated, blazar like radio
loud quasars, which account to $<5$\% of the PG quasars.  Furthermore,
these tend to be at high $\Mbh$, where $\eta$ is already uncomfortably
too high.  The host galaxy must be present at some level and would
have its greatest effect for low $L_{\rm bol}$ and high $\Mbh$ sources.
Indeed a significant host galaxy contribution is inferred by
\citet{she10} for SDSS QSOs with $L_{\rm opt} \lesssim
10^{45} \rm \; erg \; s^{-1}$.  If we adopt their correction for our
$L_{\rm opt}$, which is measured at 4861 \AA~(as opposed to 5100 \AA~in
their work) we find that several of our lowest $\eta$ sources at low
$\Mbh$ our shifted to higher values, although many still remain
with $\eta < 0.038$, the nominal lower value for maximally spinning,
counter rotating BHs.

\subsection{Effects of error in $\Mbh$}
\label{mbh}

Our $\Mbh$ estimates are described in \S\ref{mass}.  The uncertainties
on these values are somewhat difficult to estimate since they are
subject to systematic errors which are not easily quantified.  The
$M_{\rm BLR}$ estimates utilize the empirical $R_{\rm BLR}-L_{\rm
  opt}$ relation of \citet{kas05} which is calibrated against a sample
of reverberation mapping radii \citep{pet04} with 40\% intrinsic
scatter.  The reverberation mapped $\Mbh$ estimates themselves show a
scatter of 2.6-2.9 relative the $\Mbh-\sigma$ relation \citep{onk04}.
Therefore, a factor of $\sim 3$ (0.4 dex is frequently stated) is
widely reported as a characteristic uncertainty in these estimates.
However, arguments have been made that suggest the true error could be
either smaller or larger than this value.  A thorough discussion of
systematic errors, including inclination dependence, is provided by
\citet{kro01}.  Lower mass errors (some $\lesssim 0.2$ dex) on BLR
based estimates have also been suggested by some authors
\citep{kol06,she08,se10,kel10}, based primarily on the presences of
sharp features in the inferred Eddington distributions. Also, the small 
transition range of $\log \Mbh=8.5-9$, where the PG quasars transform
from radio-quiet to radio-loud, suggests a small uncertainty in $\Mbh$
\citep{lao00}.

The $\Mbh-\sigma$ estimates are rather uncertain due to difficulties
in measuring $\sigma$ in galaxies that host bright quasars.  We note
that the mean absolute deviation between the two estimates is $\Delta
\log \Mbh =0.6$ for the 13 sources with $M_\sigma$ estimates (Fig.
\ref{f:mdot}), although it is not clear which is the more reliable
estimator.

Our model fitting estimates roughly give $\eta \propto \Mbh$ at a
fixed luminosity (eq.\ref{eq:etaa}).  Therefore, errors in $\Mbh$ and
$\eta$ are correlated with $\Delta \log \Mbh \simeq \Delta \log \eta$,
scattering data points diagonally in Figure \ref{f:eta}.  Since the
inferred correlation is approximately $\eta \propto \Mbh^{1/2}$, this
scatter will have a significant projection onto the observed trend.
If we make the reasonable assumptions that mass errors are
approximately symmetric and centered on the correct $\Mbh$, we can
consider the effect they will have on our $\eta$ estimates. 

\begin{figure}[h]
\includegraphics*[width=\columnwidth]{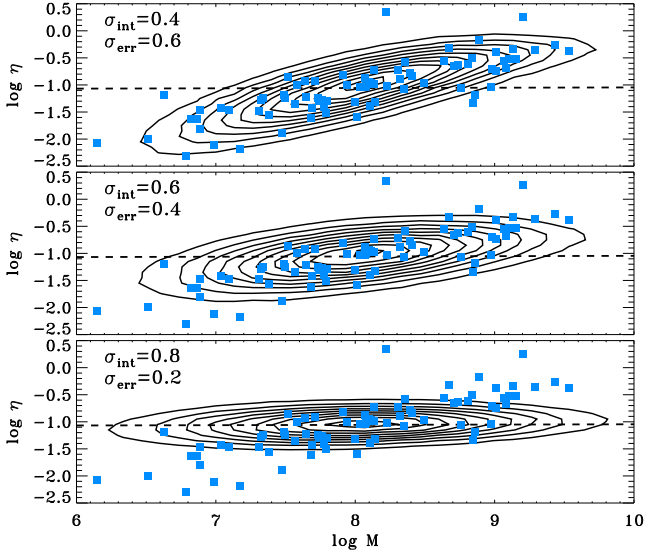}
\caption{
Simulated distributions showing the effects of mass measurement errors
on the derived $\eta$. For comparison, our measured $\eta-M$
distribution is overplotted as blue squares and the assumed mass-independent
intrinsic $\eta$ is shown as a dashed horizontal line.  The panels
show distributions with different combinations of $\sigma_{\rm int}$
and $\sigma_{\rm err}$, the standard deviations on the log normal
distributions modelling the intrinsic (true) mass distribution and the
mass measurement error, respectively.  To match the observed $\eta-\Mbh$
distribution, the PG sample needs to have both a narrow intrinsic $\Mbh$
distribution, $\sigma_{\rm int}=0.4$, and a relatively large 
measurement errors, $\sigma_{\rm err}=0.6$. With decreasing $\sigma_{\rm err}$
and increasing $\sigma_{\rm int}$ the predicted $\eta-M$ correlation
weakens (middle panel), and eventually disappears (lower panel).
\label{f:mc_eta}}
\end{figure}

In order to model the effects of mass uncertainties, we have
implemented a simple Monte Carlo procedure to generate
observed distributions of parameters based on our AD spectral models.
The procedure assumes log normal distributions for intrinsic mass
$M_{\rm int}$ and Eddington ratio $\ell = L_{\rm bol}/L_{\rm Edd}$
with means $\mu_{\rm int}$ and $\mu_{\ell}$ standard deviations
$\sigma_{\rm int}$ and $\sigma_{\ell}$, respectively.  We further
assume that the uncertainty in mass $\Delta M = M_{\rm obs} - M_{\rm
  int}$ follows a log normal distribution with a mean of zero and
standard deviation $\sigma_{\rm err}$.  Inclination $i$ is assumed to
be distributed uniformly in $\cos i$ between 0.5 and 1.  We adopt a
single, fixed value of $\eta_{\rm int}$, as our focus is on
determining whether the $\eta - M$ correlation could arise solely or
primarily from errors in our mass estimates.
 
Each Monte Carlo realization is generated by the following steps: 1)
We draw $M_{\rm int}$, $\ell$, and $i_{\rm int}$ randomly from the
above distributions.  2) We compute $L_{\rm bol,obs}= \ell M_{\rm int}
\cos i$ and $\Mdot_{\rm int} = \ell M_{\rm int}/ (\eta_{\rm int}
c^2)$.  3) We compute the observed $L_{\rm opt}$ from our AD spectral
model using $\Mdot_{\rm int}$, $M_{\rm int}$, and $i_{\rm int}$.  4)
We draw $\Delta M$ from a log normal distribution and compute $M_{\rm
  obs}=M_{\rm int} + \Delta M$. 5) We use the fitting procedure
described in \S\ref{model} (with $\cos i_{\rm obs} =0.8$) to infer an
observed accretion rate $\Mdot_{\rm obs}$ from $M_{\rm obs}$ and
$L_{\rm opt}$. 6) We infer an observed efficiency $\eta_{\rm obs}$
using equation (\ref{eq:lbol}) with $\Mdot_{\rm obs}$ and $L_{\rm
  bol,obs}$.  Note that steps 4 and 5 would simply invert steps 2 and
3 if $M_{\rm int}=M_{\rm obs}$ and $i_{\rm int}=i_{\rm obs}$.

This algorithm is repeated a large number of times to map out the
distribution of $M_{\rm obs}$ and $\eta_{\rm obs}$ for each set of
input parameters $\mu_\ell$, $\mu_M$, $\sigma_\ell$, $\sigma_{\rm
  int}$, $\sigma_{\rm err}$, and $\eta_{\rm int}$.  We focus on
distributions with $\mu_{\rm int}=\log M = 8$, $\mu_{\ell}=\log
\ell=-0.4$, $\sigma_{\ell} = 0.4$ dex, and $\eta_{\rm int}=0.08$.  The
observed distribution of $\eta$ and $M$, along with our assumption
that $\Delta M$ is log normal with zero mean, strongly constrain these
choices for $\mu_{\rm int}$ and $\eta_{\rm int}$.  We choose
$\mu_\ell$ and $\sigma_\ell$ to approximately reproduce the observed
distribution $L_{\rm bol}/L_{\rm Edd}$ seen in figure \ref{f:ledd},
although the optimal values depend slightly on $\sigma_{\rm int}$ and
$\sigma_{\rm err}$.

The resulting distributions are plotted as contours in figure
\ref{f:mc_eta}, with the observed $\eta-M$ distribution overplotted
for comparison.  Each panel shows a different combination of
$\sigma_{\rm int}$ and $\sigma_{\rm err}$.  From top to bottom in
figure \ref{f:mc_eta}, we increase $\sigma_{\rm int}$, while keeping
$\sigma_{\rm int} + \sigma_{\rm err} = 1$. This constraint is
motivated by the observed range of $M$, which is roughly consistent
with $\sigma_{\rm int} + \sigma_{\rm err} \sim 1$.

We find that we can largely reproduce the observed $\eta-M$
correlation by choosing a sufficiently small $\sigma_{\rm int}$ and a
sufficiently large $\sigma_{\rm err}$, although difficulties remain
for this interpretation.  As noted above, $\sigma_{\rm err} \sim 0.4$
dex is often asserted, but this value would be insufficient to
reproduce all of the correlation by mass errors alone.  Even if one
allows that $\sigma_{\rm err}=0.6$ dex and $\sigma_{\rm int}=0.4$ dex
are theoretically viable, this distribution has mixed results
simultaneously reproducing all of the observables.

\begin{figure}[h]
\includegraphics*[width=\columnwidth]{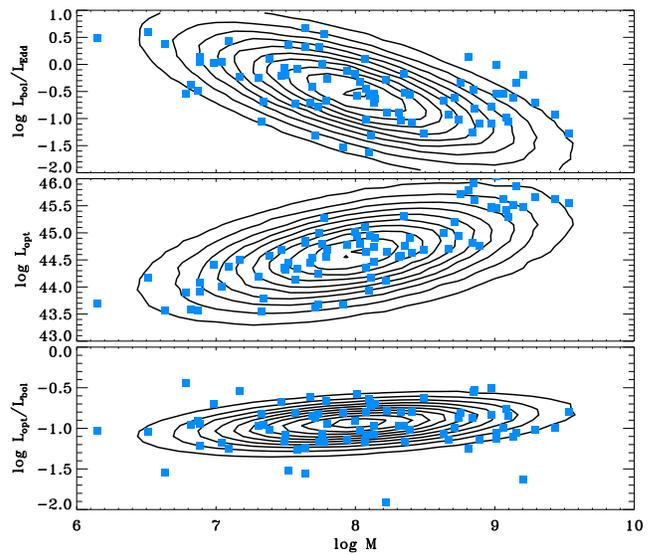}
\caption{
Simulated distributions showing the effects of mass measurement errors
on the observed distributions of $L_{\rm bol}/L_{\rm Edd}$, $L_{\rm
  opt}$, and $L_{\rm opt}/L_{\rm bol}$. For comparison, observables
from our PG sample are overplotted as blue squares.  The distributions
are computed assuming $\sigma_{\rm err}=0.6$ dex and $\sigma_{\rm
  int}=0.4$ dex, corresponding to the $\eta-M$ distribution plotted
in the top panel of figure \ref{f:mc_eta}. Note that although the observed 
distributions in the upper and lower panels can be reproduced, the 
observed $L_{\rm opt}$ vs. $\Mbh$ correlation is stronger than expected
if the error in $\Mbh$ is as large as required.
\label{f:mc_vars}}
\end{figure}

This can be seen in figure \ref{f:mc_vars}, which shows the two
dimensional distributions of $M_{\rm obs}$ with $L_{\rm bol}/L_{\rm
  Edd}$, $L_{\rm opt}$, and $L_{\rm opt}/L_{\rm bol}$ for the
$\sigma_{\rm err}=0.6$ dex and $\sigma_{\rm int}=0.4$ dex
distribution.  Even though our intrinsic distribution of $L_{\rm
  bol}/L_{\rm Edd}$ is mass independent, an anti-correlation with
$\Mbh$ arises which is qualitatively consistent with the observed
distribution.  Reasonable agreement is also found for $L_{\rm
  opt}/L_{\rm bol}$.  However a significant discrepancy remains
between the simulated and observed $L_{\rm opt}$ distributions, which
shows a much tighter correlation of $L_{\rm opt}$ with $\Mbh$ in the
observed data.  The breadth of the simulated distribution is a result
of the scatter induced by mass errors.  We can get a narrower, more
correlated distribution by reducing $\sigma_{\rm err}$ to 0.2 dex, but
this produces a steeper dependence of $L_{\rm opt}$ on $\Mbh$ and
doesn't produces a large enough $\eta - \Mbh$ correlation.

It is plausible that some of remaining discrepancy could be addressed
with more complex distributions for the $M_{\rm int}$, $\ell$, and
$\Delta M$.  We have no strong empirical or theoretical motivation to
believe that the real distributions of the above variables are log
normal.  In fact, these assumptions lead to an $M_{\rm obs}$
distribution which is also log normal.  This conflicts slightly with
our observed $M$ distribution, which is somewhat ``flatter'' (i.e. has
negative kurtosis) than a log normal distribution with the same
variance.  We have little empirical knowledge of the true
distribution, which is subject to the non-trivial selection effects of
the PG sample.  Therefore, we can only hope to approximately quantify
the effects of $M$ uncertainties and we adopt the above assumptions
largely for the sake of simplicity.  A more sophisticated analysis,
along the lines of \citet{kel10} is beyond the scope of the present
work and possibly not amenable to our limited sample size.

Regardless of the precise details, this analysis relies on the
(plausible) assumption of an approximately symmetric mass measurement
error and a centrally peaked intrinsic mass distribution.  At high
mass, such a distribution could arise because large $\Mbh$ objects are
rare, but the turnover at low $\Mbh$ would require selection effects
that tend to exclude low $\Mbh$ objects.  This could be the case if
low $\Mbh$ systems are Eddington limited and tend to have too large of
a host fraction to make it into our PG sample.

We note that our treatment of inclination has little effect on the
resulting distributions.  We model a uniform distribution of $\cos i$
from 0.5-1, but assume a single value of $\cos i=0.8$ for our observed
inclination.  In the absence of mass errors, this introduces a modest
broadening of $\eta_{\rm obs}$ about $\eta_{\rm int}$, and the
magnitude of the effect is consistent with our above estimates in \S
\ref{lbolerr}.  However, this is dwarfed by the broadening
introduced by any plausible mass error (i.e. any $\sigma_{\rm err} \ge
0.2$ dex).  A dependence on inclination may also be present in our
virial $\Mbh$ estimates, since the translation of the line width to a
virial velocity will in general depend on the inclination distribution
of the emitting gas, which may (in turn) be related to the
inclination of the AD.  Therefore, inclination uncertainties may have
a much stronger impact through their effect on $\Mbh$ than their
effect on either $L_{\rm opt}$ or $L_{\rm bol}$.  However, we consider
this effect implicitly modelled by our mass error analysis above. 

In addition to the distributions shown in figures \ref{f:mc_eta} and
\ref{f:mc_vars}, which do not model dust reddening, we have generated
Monte Carlo simulations that incorporate the effects of reddening on
the observed $L_{\rm bol}$ and $L_{\rm opt}$.  We consider several
different reddening curves, including SMC-like \citep{ric03} and those
of \citet{cze04} and \citet{gas04}. As expected, we find that
reddening tends to lower the ratio of $L_{\rm bol}/L_{\rm opt}$,
decreasing the inferred $\eta$.  This result is independent of
reddening curve, although the amount of extinction required depends on
the curve used.  We find moderate amounts of extinction ($\rm{E(B-V)}
\lesssim 0.1$ for SMC-like reddening) are all that is needed to explain
some of low $\eta$ sources that populated the lower envelope of the
observed distribution.  The decrease in $\eta$ tends to be slightly
larger for low $\Mbh$, which have a larger fraction (relative to high
$\Mbh$ systems) of their bolometric output in the UV in our model
SEDs.  This effect alone introduces a slight positive correlation
between $\eta$ and $M$, but the magnitude of slope change is small and
cannot contribute significantly to the observed trend for reasonable
extinction values.

Our conclusions from this analysis is that mass measurement errors can
introduce a spurious correlation of $\eta$ with $\Mbh$, and that this
effect almost certainly contributes to some of the inferred
correlation.  However, given the tight correlation between $L_{\rm
  opt}$ and $\Mbh$, and the requirement of a very narrow mass
distribution ($\sigma_{\rm int} \lesssim 0.4$ dex), we disfavor $\Mbh$
errors alone as an explanation for the observed correlation.

\subsection{The accretion disk model}

Since our $\eta$ estimates are all based on the bare, radiatively
efficient thin disk model \citep{ss73,nt73}, it is conceivable that
the evolution in $\eta$ with $\Mbh$ is a result of inapplicability of
the underlying spectral model.  Indeed, it is well known that this
model has difficulties reproducing AGN observations.  \citet{kb99}
offer a thorough review of discrepancies between models and
observations.  In particular, it fails to account for the X-ray
emission and tends to predict spectral slopes which rise too steeply
in the UV \citep[see e.g.][ and the AD model to SED comparison for the
low $\Mbh$ objects in Figure \ref{f:seds}]{ant99,bon07,dwb07}.
However, the model has had some success in other sources.  It does a
rather good job of reproducing the thermally dominant states of BH
X-ray binaries, in which the emission is believed to come
predominantly from an AD.  Detailed spectral models similar to the
ones used here \citep{dh06} not only reproduce the spectrum, but also
the spectral evolution as $\Mdot$ varies \citep[e.g][]{ddb06,sha06}.

A central premise to our $\eta$ estimates is that some mechanism acts
to redistribute flux from UV to the X-ray frequencies.  In principle,
such a mechanism can account for the mismatch between observation and
theory in the UV while simultaneously explaining the larger than
expected X-ray flux.  This could be related to advection, a
Comptonizing coronae, a warm skin, etc.  Regardless of the mechanism
invoked, our $\eta$ estimates rely on the assumption that it does not
act in regions where most of the optical photons are emitted.  If some
mechanism is acting on the optically emitting regions which breaks the
assumption that local radiation flux (and work done by the accretion
stress) balances the local release of binding energy, it could clearly
modify our estimates.

This standard disk model gives, at best, mixed results for explaining
the sizes of continuum emission regions inferred from microlensing
constraints in gravitationally lensed QSOs.  The major discrepancy is
that the optical to UV emission is constrained to come from radii
which are factors of $\sim3-10$ larger than expected from a standard
disk \citep[e.g.][]{mor05,poo06,poo07,mor10,mor08,dai10}.  It also
tends to give $\eta$ estimates which are lower ($\log \eta =
-1.77+\log l$) than those derived from the Soltan argument
\citep{mor10}.  More optimistically, the relative scaling of radius
with observing wavelength ($R \propto \lambda^{4/3}$) is consistent
with microlensing constraints in several sources
\citep[e.g.][]{eig08,pmk08}.  Furthermore, the $\Mbh$ dependence of
the microlensing radii \citep{mor10} are also consistent with
theoretical expectations ($R \propto \Mbh^{2/3}$).

In principle, one could reconcile the discrepancy in absolute
microlensing radii by assuming a flux profile which falls off less
steeply with radius.  This would also improve the agreement between
observations and spectral slopes.  If we parametrize the radiative
flux as $F \propto R^{-\beta}$, the standard disk corresponds to
$\beta=3$ and the same arguments which leads to equation
(\ref{eq:lnu2}) yields
\begin{equation}
\nu L_\nu \propto \Mdot^{2/\beta} \Mbh^{(2-4/\beta)} \nu^{(4-8/\beta)}
\label{eq:beta}
\end{equation}
As $\beta$ decreases the spectral slope flattens, with $\nu L_\nu$
independent of $\nu$ at frequencies that are unaffected by truncation
at the inner or outer radii of the disk.  Therefore, a model with
$\beta < 3$ would give larger radii for the optical emitting regions
and also flatter optical to UV spectral slopes, in better agreement
with observations.

Suggestively, a model with $\beta$ closer to 2 would also give a ratio
of $L_{\rm opt}/L_{\rm bol}$ which is less dependent on $\Mbh$ at
fixed $\eta$.  This can be inferred from equation (\ref{eq:beta}),
which shows that the dependence of $\nu L_\nu$ on $\Mbh$ weakens as
$\beta \rightarrow 2$. For $\beta = 2$, the $\Mdot$ needed to match
$L_{\rm opt}$ would be independent of $\Mbh$ and a constant $\eta$
model would be consistent with a constant $L_{\rm opt}/L_{\rm bol}$.
Such a scaling could, for example, result from irradiation.  If the
optical emission came from a flared region of the outer disk which was
irradiated by the emission from the inner disk, the reprocessed
radiation could have a radial profile with $\beta$ closer to 2 at
larger radii.  Thus one might obtain a more constant $\eta$ by
accounting for irradiation in the model if it dominates the intrinsic,
local emission.

However, there are two main problems with such an interpretation.
First, it would give a scaling $R \propto \lambda^{4/\beta}$ that
would not agree with microlensing constraints, which as noted above
are consistent with $\beta \sim 3$ \citep{eig08,pmk08}.  Secondly, if
the sources are irradiated, we must be observing more optical emission
than we would have from a bare AD.  By using $L_{\rm opt}$ without
accounting for the ``extra'' reprocessed emission we would have
overestimated $\Mdot$ and underestimated $\eta$.  Although irradiation
could then account for the low $\Mbh$ systems that have low $\eta$, it
could only increase our already uncomfortably high $\eta$ estimates
for large $\Mbh$.  Therefore, we do not believe that irradiation alone
can explain the $\eta - \Mbh$ correlation.

If the assumed redistribution mechanism that shifts emission from the
UV to X-ray depends mainly on $r=R/R_g$, it may be the case that the
optical emission will be increasingly influenced by the mechanism as
$\Mbh$ increases.  This is plausible if the redistribution mechanism
primarily acts in a confined region near the innermost radius of the
disk, such as might be the case for Comptonization of disk photons in
a compact corona. As $\Mbh$ increases the range of optically emitting
radii moves to smaller $r$.  Therefore a larger fraction of the
optical emission may be subject to the redistribution, and the
observed $L_{\rm opt}$ may be underestimating the true $\Mdot$ and
overestimating $\eta$.  Such a scenario could potentially explain the
high values of $\eta$ at large $\Mbh$, but cannot account for the low
$\eta$ at small $\Mbh$.

Another possibility is that the assumption of a constant $\Mdot$
through the disk may be incorrect if there is outflow interior to the
optical emitting region that carries away a sizable fraction of the
accreting mass. Such an outflow converts some of the accretion power
into mechanical luminosity, rather than radiative luminosity.  The
outflow can be in the form of a wind, as seen in broad absorption line
quasars, or as a jet, seen in radio loud quasars. A high mechanical
luminosity can be accommodated in low $\Mbh$ AGN, where the radiative
$\eta$ is low. However, low $\Mbh$ and low $L_{\rm opt}$ AGN generally
do not show high velocity outflows, in particular in the PG sample
\citep{lb02}. In high $\Mbh$ systems only a small fraction of the
accretion power can go out as mechanical luminosity, as the radiative
efficiency is close to maximal. However, both jets and fast UV
absorbing outflows are commonly seen in high $\Mbh$ and high $L_{\rm
  opt}$ systems, suggesting they carry only a small fraction of the
radiative power. Thus, mechanical outflows does not appear to go in
the direction that can weaken the correlation of $\eta$ with
$\Mbh$. But, the high $\eta$ values in high $\Mbh$ systems provides an
interesting limit on the possible mechanical feedback of AGN on their
environment. We note in passing that some of the AGN radiative and
mechanical luminosity may be obtained by tapping the BH spin, which
can drive $\eta$ to values above the theoretical AD upper limit.

\subsection{Evolution of $\eta$ with $\Mbh$}

Thus far we have focused on potential errors in our assumptions which
could artificially introduce correlations between $\eta$ and $\Mbh$,
but it is entirely possible that such a correlation is real.  As noted
previously, our mean $\log \eta$ is consistent with estimates derived
with the Soltan argument and the range of inferred $\eta$ is
physically realizable. Given the uncertainties in $\Mbh$ and $L_{\rm
  bol}$, it is entirely plausible that all $\eta$ estimates are below
the theoretical maximum for a rapidly accreting BH
($a_*=0.998$) with no torque on the inner boundary.  If
magnetohydrodynamic torques are present, the spin may be
limited to lower values ($a_* \sim 0.9$, \citealt{gsm04},
and references therein).  However, torques at the inner edge of the
disk could, in principle, yield even higher efficiencies \citep{ak00}.

If $a_*$ is the predominant factor which determines $\eta$, as in the
standard thin disk model \citep{ss73,nt73}, one would expect $\eta$ to
vary with $\Mbh$ if $a_*$ varies with $\Mbh$.  In fact, detailed
semi-analytic models of galaxy formation that simulate the accretion
histories of super-massive BHs can find a strong dependence of $a_*$
on $\Mbh$ \citep{lpc09,fan09}.  For their Model A, \citet{lpc09} find
a general trend of the average $a_*$ increasing with $\Mbh$.  For
$\Mbh \lesssim 10^6 \Msun$, they find a range of $a_*$, covering the
range from 0 to 1, but with more systems at low $a_*$.  At $\Mbh
\gtrsim 10^8 \Msun$, they find most systems with $a_* \sim 1$.  For
their chaotic accretion model, \citet{fan09} mostly find $a_* \lesssim
0.5$ for $\Mbh \lesssim 10^8 \Msun$, with some systems counter
rotating ($a_* \lesssim 0$).  At higher mass $a_*$ increases with a
median $a_* \sim 0.6-0.8$ for $\Mbh \gtrsim 10^9 \Msun$.  The
prolonged accretion model of \citet{fan09} yields a much different
distribution, with $a_* \sim 1$ for $\Mbh \lesssim 10^8 \Msun$ and
dropping to $a_* \sim 0.8-0.9$ at higher mass.

\begin{figure}
\includegraphics*[width=\columnwidth]{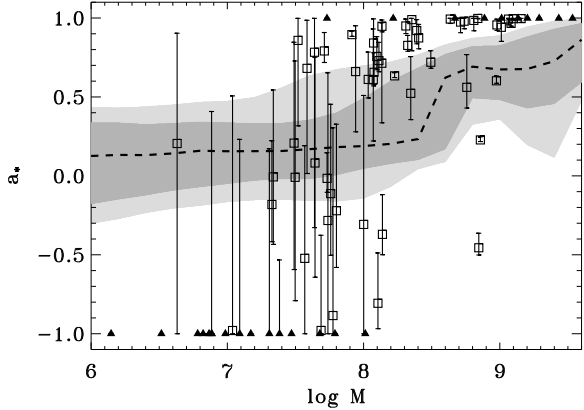}
\caption{
Spins as a function of $\Mbh$ for the 80 PG QSOs in our sample.  We
use the the \citet{nt73} model to derive $a_*$ for each value of
$\eta$.  Values of $\eta > 0.31$ or $\eta < 0.038$ are set to these
limits, corresponding to $a_*=0.998$ and $a_*=-1$, respectively, and
represented by filled triangles for plotting purposes. The massive BHs
with $\log M>8$ generally have $a_*>0.5$, and the lower mass BHs have
$a_*<0.5$.  For comparison we plot contours representing the
\citet{fan09} $a_*$ distribution (see their figure 9).  The thick
dashed curve represents the median, while the shaded contours contain
60\% and 80\% of systems at the corresponding mass. This is broadly
consistent with various earlier suggestions for a rise in $a_*$ with
$M$.
\label{f:spins}}
\end{figure}

In order to compare with these and other predictions, we convert
$\eta$ to $a_*$ for each source in our sample.  This is done assuming
disk model with no inner torque and an inner radius corresponding to
$r_{\rm ms}$.  The minimum and maximum $\eta$ correspond to 0.42 and
0.038 for models with $a_*=1$ and -1, respectively.  The results are
plotted in figure \ref{f:spins}.  In cases where our $\eta$ estimates
lie outside the allowed range, we assume $a_*=-1$ or 1 for $\eta$
below and above the allowed range, respectively.  The relatively
gradual evolution in $\eta$ with $\Mbh$ translates into a more abrupt
evolution in $a_*$ due to the rapid rise in $\eta$ as $a_*$ approaches
unity.  For comparison purposes, we have plotted dashed curves
outlining the $a_*$ distribution of the \citet{fan09} chaotic
accretion model in figure \ref{f:spins}.  The thick curve is the
median of the distribution and the shaded contours contain 60\% and
80\% of systems at the corresponding mass.

Both model A of \citet{lpc09} and the chaotic accretion model of
\citet{fan09} are qualitatively consistent with our results, in that
the average $a_*$ increases with $\Mbh$.  However, neither is a
perfect match to our correlation.  Our estimates would not accommodate
the high $a_*$ at low $\Mbh$ in \citet{lpc09}, and our high $\Mbh$
systems would require $a_*$ very near unity, so the spins in this
range in the \citet{fan09} model are not quite high enough.  Of
course, both of these calculations are subject to uncertainties in
what they assume about the angular momentum of the accreting gas, so
there may be some leeway to accommodate our results.  The prolonged
accretion model of \citet{fan09} would be completely inconsistent with
our estimates.

Further support for spin evolution comes from models for radio
emission in QSOs that require high $a_*$ (e.g \citealt{bz77}).  Since
radio loud QSOs are associated with large $\Mbh$, high $a_*$ would be
required at large $\Mbh$ if these models are correct.  The minimum
value of $a_*$ needed for significant jet production is somewhat
uncertain in these models, so it is difficult to make this a
quantitative argument, but it is qualitatively consistent with our
finding of high $\eta$ at large $\Mbh$.

Other factors besides $a_*$ may also play a role in setting $\eta$.
One possibility is that advection may be operating near the inner-most
radii for systems accreting near the Eddington rate \citep{abr88}.
Figure \ref{f:ledd} shows our $L_{\rm bol}/L_{\rm Edd}$ distribution
as a function of $\Mbh$.  Most of the systems with $L_{\rm bol}/L_{\rm
  Edd} > 1$ have $\Mbh < 10^8 \Msun$.  This may, in part, be due to
the lowest $\Mbh$ systems having their $\Mbh$ predominantly
underestimated as discussed above.  Alternatively, it may be a
selection effect resulting from the requirement that the QSO outshine
its host galaxy, which will require a higher $L_{\rm bol}/L_{\rm Edd}$
in disk dominated galaxies.  If these systems are truly accreting at
or above the Eddington rate, the advection time is comparable to the
photon diffusion time and a substantial fraction of the radiation
might get advected into the BH before radiating.  If these sources
are accreting at only mildly super-Eddington rates, advection should
have no effect at larger radii where the optical is emitted, and our
$\Mdot$ estimates should still be reliable.  These systems would then
be truly radiatively inefficient.  Comparison of figures \ref{f:eta}
and \ref{f:ledd} show that the effect of advection in flows with
$L_{\rm bol}/L_{\rm Edd} \gtrsim 1$ could plausibly account for some,
but not all, of the lowest $\eta$ sources.

\section{Conclusions}

We have estimated the accretion rate by fitting radiatively efficient
AD model SEDs (hereafter the standard model) to the optical emission
in a sample of 80 PG QSOs.  This method is insensitive to properties
of the accretion flow and BH spacetime near the inner edge of the disk
(torques at the inner radius, BH spin, advection, etc.) as long as the
emission comes from large radius.  We use detailed AD model SEDs which
are computed from non-LTE atmospheres and include relativistic effects
on photon geodesics, but find our results are qualitatively reproduced
by simple, non-relativistic local blackbody relations. The
derived accretion rates are nearly insensitive to spin at low black
hole mass ($\Mbh \lesssim 10^8 \Msun$) and only weakly sensitive at
higher mass when the fitting is done at optical frequencies.  The
accretion rates are more sensitive to assumed BH masses, which are
estimated using broad line region virial methods as well as masses
derived from the $\Mbh-\sigma$ relation for 13 sources with bulge
velocity dispersion measurements.

Our sample of 80 PG QSOs was chosen because they have ample broadband
coverage at optical to far UV and X-ray frequencies.  This allowed us
to robustly estimate the bolometric luminosity modulo some uncertainty
in the unobservable EUV emission.  These luminosities, combined with
our estimates of the accretion rate, allow us to compute the radiative
efficiency for each source in our sample.  We find a mean efficiency
of $\log \eta = -1 \pm 0.5$, in agreement with integral constraints
derived by matching local black hole mass density to the integrated
quasar luminosity function (i.e. the Soltan argument).  This basic
agreement suggests that the standard model can provide a reasonable
first-order approximation to real accretion flows, at least at radii
where most of the optical emission is produced.

We find a strong correlation of efficiency with BH mass (approximately
$\eta \propto \Mbh^{1/2}$ in our sample) extending from $\eta \sim
0.01$ for low masses to to $\eta \lesssim 1$ at the highest masses.
This relation arises because the ratio of the optical to bolometric
luminosity is roughly independent of BH mass, whereas a constant
efficiency thin disk model would predict a substantial increase in
this ratio as mass increases.

We consider three possibilities for explaining the $\eta - \Mbh$
correlation:\\
 
\noindent
1) The correlation is real.  It could plausibly arise from a mass
dependence of the BH spin driven by the differing accretion histories
of black holes at different masses.  Semi-analytic models of galaxy
formations that attempt to model spin distributions of supermassive
BHs (e.g. \citealt{lpc09,fan09}) find spin dependencies which are
qualitatively, although not quantitatively, consistent with our
trend of increasing efficiency with increasing BH mass. \\

\noindent
2) One or more of the observables or input parameters to the model are
incorrectly estimated.  Indeed, scatter in the broad line region based
BH mass estimates probably contributes to some of the trend.  We argue
that the highest and lowest BH masses will tend to be overestimated
and underestimated, respectively.  This, in turn, leads to
overestimates of the efficiency at the highest masses and
underestimates at lowest masses.  However, it would be difficult for
these correlated errors to explain all of the trend, unless the true
mass distribution is very narrow ($\lesssim 1$ dex), which seem highly
unlikely given the large range of bolometric luminosities in our
sample and the significantly larger range of $\Mbh$ derived in other
studies of AGN.\\

\noindent
3) The standard (bare) AD model does not adequately approximate the
dependence of real accretion flows on mass.  It has well-known
difficulties reproducing the observed SEDs of AGN and microlensing
sizes of emission regions.  A central assumption of this work is that
standard model works well at larger radii where the optical emission
is predominantly produced, but that some mechanism operates very near
the black hole to redistribute flux from UV to X-ray frequencies.  It
is possible that the standard model fails in the optical emitting regions
as well and that real flows naturally give rise to a relatively
constant ratio of optical to bolometric luminosity which does not
depend strongly on mass.  We consider a number of modifications to the
standard model, including irradiation.  However, no single possibility
we considered seemed capable of explaining the whole trend at both low
and high masses, while maintaining a nearly constant $\eta$ whose
value was consistent with Soltan argument.\\

The role of errors in the mass estimates could be definitively
addressed by more precise mass estimates.  This would either require
reducing the scatter in the BLR estimates or obtaining a larger sample
of sources with well-measured bulge velocity dispersions.
Differentiating the degree to which the trend reflects real evolution
in efficiency with mass or inapplicability of the standard models
assumptions is more difficult.  Microlensing models can provide
independent estimates for the efficiency in lensed QSOs \citep{mor10},
although only a handful are currently available.  Alternatively,
relativistically broadened Fe K$\alpha$ lines could provide spin
estimates from which we could infer efficiencies, but (again) precise
constraints are only available for a few sources.  Some light on what
controls $\eta$ may be shed by the variability of $\eta$ on timescales 
longer than the viscous timescales, when the AD may be quasistatic.
If $L_{\rm bol}\propto L_{\rm opt}^{1.5}$, then $\eta$ is constant,
which is consistent with being spin driven. Until such
studies are available for a significant
sample of sources, this will likely remain an open question.

\acknowledgements{We thank the referee, Chris Reynolds, for
  suggestions that significantly improved our manuscript.  We further
  thank Omer Blaes, Julian Krolik, Scott Tremaine, and Glenn van de
  Ven for helpful discussions.  We thank Nikos Fanidakis for helpful
  discussions and for providing his data, which we used in Figure
  \ref{f:spins}. We also thank Alexei Baskin for providing the UV
  spectral slopes. SWD acknowledges support through grants NSF
  AST-0807432, NASA NNX08AH24G, NSF AST-0807444, and NASA grant number
  PF6-70045, awarded by the Chandra X-ray Center, which is operated by
  the Smithsonian Astrophysical Observatory for NASA under contract
  NAS8-03060. SWD is currently supported in part by NSERC of
  Canada. AL acknowledges the hospitality and support of the Institute
  for Advanced Study, where this research was initiated."}

\clearpage

\LongTables

\begin{deluxetable}{lccccccc}
\tablecolumns{8}
\tablewidth{0pt}
\tablehead{
\colhead{Object} &
\colhead{H$\beta$ FWHM\tablenotemark{a}} &
\colhead{$\log L_{\rm opt}$\tablenotemark{b}} &
\colhead{$\log M_{\rm BLR}$\tablenotemark{c}} &
\colhead{$\log\Mdot$\tablenotemark{d}} &
\colhead{$\alpha_{\rm uv}$} &
\colhead{$\log L_{\rm bol}$\tablenotemark{b}} &
\colhead{$\log \eta$}
}
\startdata
$0003+158$ &
$4760$ &
$45.87$ &
$ 9.16$ &
$  0.79$ &
$ -0.22 \pm  0.04$ &
$ 46.92 \pm  0.25$ &
$-0.52$ \\
$0003+199$ &
$1640$ &
$43.91$ &
$ 6.88$ &
$ -0.06$ &
$  0.14 \pm  0.04$ &
$ 45.13 \pm  0.35$ &
$-1.47$ \\
$0007+106$ &
$5100$ &
$44.55$ &
$ 8.31$ &
$ -0.42$ &
$ -0.60 \pm  0.23$ &
$ 45.52 \pm  0.17$ &
$-0.72$ \\
$0026+129$ &
$1860$ &
$44.99$ &
$ 7.74$ &
$  0.80$ &
$ -0.14 \pm  0.07$ &
$ 46.15 \pm  0.29$ &
$-1.30$ \\
$0043+039$ &
$5300$ &
$45.47$ &
$ 8.98$ &
$  0.36$ &
$ -2.61 \pm  0.06$ &
$ 45.98 \pm  0.02$ &
$-1.04$ \\
$0049+171$ &
$5250$ &
$43.68$ &
$ 7.73$ &
$ -1.19$ &
$  4.07 \pm  1.05$ &
$ 48.03 \pm  0.56$ &
$ 2.56$ \\
$0050+124$ &
$1240$ &
$44.41$ &
$ 6.99$ &
$  0.58$ &
$ -2.46 \pm  0.02$ &
$ 45.12 \pm  0.04$ &
$-2.12$ \\
$0052+251$ &
$5200$ &
$45.00$ &
$ 8.64$ &
$ -0.04$ &
$ -1.08 \pm  0.06$ &
$ 46.06 \pm  0.25$ &
$-0.55$ \\
$0157+001$ &
$2460$ &
$45.02$ &
$ 8.00$ &
$  0.59$ &
$ -0.45 \pm  0.28$ &
$ 45.93 \pm  0.23$ &
$-1.31$ \\
$0804+761$ &
$3070$ &
$44.79$ &
$ 8.03$ &
$  0.20$ &
$ -1.53 \pm  0.13$ &
$ 45.82 \pm  0.11$ &
$-1.03$ \\
$0838+770$ &
$2790$ &
$44.56$ &
$ 7.79$ &
$  0.08$ &
$ -2.43 \pm  0.38$ &
$ 45.22 \pm  0.02$ &
$-1.51$ \\
$0844+349$ &
$2420$ &
$44.31$ &
$ 7.50$ &
$ -0.01$ &
$ -1.27 \pm  0.24$ &
$ 45.40 \pm  0.28$ &
$-1.24$ \\
$0921+525$ &
$2120$ &
$43.56$ &
$ 6.87$ &
$ -0.55$ &
$ -0.85 \pm  0.20$ &
$ 44.47 \pm  0.14$ &
$-1.64$ \\
$0923+129$ &
$1990$ &
$43.58$ &
$ 6.82$ &
$ -0.49$ &
$ -0.79 \pm  0.28$ &
$ 44.53 \pm  0.15$ &
$-1.63$ \\
$0923+201$ &
$7610$ &
$44.81$ &
$ 8.84$ &
$ -0.47$ &
$ -1.43 \pm  0.46$ &
$ 45.68 \pm  0.05$ &
$-0.51$ \\
$0947+396$ &
$4830$ &
$45.20$ &
$ 8.71$ &
$  0.19$ &
$ -1.19 \pm  0.05$ &
$ 46.20 \pm  0.28$ &
$-0.65$ \\
$1001+054$ &
$1740$ &
$44.69$ &
$ 7.47$ &
$  0.59$ &
$ -1.13 \pm  0.07$ &
$ 45.36 \pm  0.12$ &
$-1.88$ \\
$1011-040$ &
$1440$ &
$44.08$ &
$ 6.89$ &
$  0.17$ &
$ -0.34 \pm  0.20$ &
$ 45.02 \pm  0.23$ &
$-1.81$ \\
$1012+008$ &
$2640$ &
$44.95$ &
$ 8.01$ &
$  0.46$ &
$ -2.32 \pm  0.48$ &
$ 45.53 \pm  0.02$ &
$-1.59$ \\
$1022+519$ &
$1620$ &
$43.56$ &
$ 6.63$ &
$ -0.36$ &
$  0.62 \pm  0.43$ &
$ 45.10 \pm  0.39$ &
$-1.19$ \\
$1048-090$ &
$5620$ &
$45.45$ &
$ 9.01$ &
$  0.30$ &
$  0.03 \pm  1.59$ &
$ 46.57 \pm  0.32$ &
$-0.38$ \\
$1048+342$ &
$3600$ &
$44.74$ &
$ 8.13$ &
$  0.02$ &
$ -8.18 \pm  3.80$ &
$ 45.70 \pm  0.31$ &
$-0.98$ \\
$1049-006$ &
$5360$ &
$45.46$ &
$ 8.98$ &
$  0.34$ &
$ -0.83 \pm  0.06$ &
$ 46.29 \pm  0.15$ &
$-0.71$ \\
$1100+772$ &
$6160$ &
$45.51$ &
$ 9.13$ &
$  0.29$ &
$ -1.76 \pm  0.06$ &
$ 46.61 \pm  0.25$ &
$-0.34$ \\
$1103-006$ &
$6190$ &
$45.43$ &
$ 9.08$ &
$  0.21$ &
$ -1.18 \pm  0.09$ &
$ 46.19 \pm  0.10$ &
$-0.68$ \\
$1114+445$ &
$4570$ &
$44.75$ &
$ 8.35$ &
$ -0.16$ &
$ -1.79 \pm  0.04$ &
$ 45.92 \pm  0.49$ &
$-0.58$ \\
$1115+407$ &
$1720$ &
$44.58$ &
$ 7.38$ &
$  0.49$ &
$ -0.62 \pm  0.07$ &
$ 45.59 \pm  0.21$ &
$-1.56$ \\
$1116+215$ &
$2920$ &
$45.31$ &
$ 8.35$ &
$  0.69$ &
$ -0.67 \pm  0.03$ &
$ 46.27 \pm  0.13$ &
$-1.08$ \\
$1119+120$ &
$1820$ &
$44.01$ &
$ 7.04$ &
$ -0.06$ &
$ -0.24 \pm  0.18$ &
$ 45.18 \pm  0.34$ &
$-1.42$ \\
$1121+422$ &
$2220$ &
$44.80$ &
$ 7.76$ &
$  0.48$ &
$ -0.48 \pm  0.07$ &
$ 45.87 \pm  0.16$ &
$-1.27$ \\
$1126-041$ &
$2150$ &
$44.19$ &
$ 7.31$ &
$ -0.02$ &
$ -1.46 \pm  0.27$ &
$ 45.16 \pm  0.28$ &
$-1.47$ \\
$1149-110$ &
$3060$ &
$43.79$ &
$ 7.34$ &
$ -0.66$ &
$ -0.62 \pm  0.49$ &
$ 44.75 \pm  0.18$ &
$-1.24$ \\
$1151+117$ &
$4300$ &
$44.65$ &
$ 8.23$ &
$ -0.20$ &
$ -2.65 \pm  0.51$ &
$ 45.43 \pm  0.01$ &
$-1.02$ \\
$1202+281$ &
$5050$ &
$44.58$ &
$ 8.32$ &
$ -0.38$ &
$ -1.45 \pm  0.11$ &
$ 45.39 \pm  0.07$ &
$-0.89$ \\
$1211+143$ &
$1860$ &
$44.85$ &
$ 7.64$ &
$  0.68$ &
$ -0.76 \pm  0.07$ &
$ 46.41 \pm  0.50$ &
$-0.93$ \\
$1216+069$ &
$5190$ &
$45.62$ &
$ 9.06$ &
$  0.51$ &
$ -0.97 \pm  0.03$ &
$ 46.61 \pm  0.28$ &
$-0.55$ \\
$1226+023$ &
$3520$ &
$46.03$ &
$ 9.01$ &
$  1.18$ &
$ -0.64 \pm  0.01$ &
$ 47.09 \pm  0.24$ &
$-0.74$ \\
$1229+204$ &
$3360$ &
$44.24$ &
$ 7.73$ &
$ -0.35$ &
$ -0.98 \pm  0.10$ &
$ 45.06 \pm  0.04$ &
$-1.25$ \\
$1244+026$ &
$ 830$ &
$43.70$ &
$ 6.15$ &
$  0.15$ &
$ -0.43 \pm  0.32$ &
$ 44.74 \pm  0.22$ &
$-2.07$ \\
$1259+593$ &
$3390$ &
$45.79$ &
$ 8.81$ &
$  0.99$ &
$ -0.74 \pm  0.04$ &
$ 47.04 \pm  0.29$ &
$-0.61$ \\
$1302-102$ &
$3400$ &
$45.71$ &
$ 8.76$ &
$  0.92$ &
$ -1.85 \pm  0.04$ &
$ 46.51 \pm  0.12$ &
$-1.06$ \\
$1307+085$ &
$4190$ &
$44.92$ &
$ 8.39$ &
$  0.05$ &
$ -1.16 \pm  0.07$ &
$ 45.93 \pm  0.20$ &
$-0.78$ \\
$1309+355$ &
$2940$ &
$44.95$ &
$ 8.11$ &
$  0.37$ &
$ -1.66 \pm  0.07$ &
$ 45.63 \pm  0.05$ &
$-1.40$ \\
$1310-108$ &
$3630$ &
$43.56$ &
$ 7.33$ &
$ -1.00$ &
$ -1.13 \pm  0.44$ &
$ 44.37 \pm  0.10$ &
$-1.28$ \\
$1322+659$ &
$2790$ &
$44.78$ &
$ 7.94$ &
$  0.27$ &
$ -0.67 \pm  0.05$ &
$ 45.92 \pm  0.29$ &
$-1.01$ \\
$1341+258$ &
$3040$ &
$44.13$ &
$ 7.57$ &
$ -0.37$ &
$ -0.85 \pm  0.75$ &
$ 44.94 \pm  0.16$ &
$-1.35$ \\
$1351+236$ &
$6540$ &
$43.93$ &
$ 8.10$ &
$ -1.14$ &
$ -1.20 \pm  0.90$ &
$ 44.57 \pm  0.12$ &
$-0.95$ \\
$1351+640$ &
$5660$ &
$44.69$ &
$ 8.49$ &
$ -0.38$ &
$ -1.70 \pm  0.05$ &
$ 45.31 \pm  0.05$ &
$-0.97$ \\
$1352+183$ &
$3600$ &
$44.65$ &
$ 8.07$ &
$ -0.06$ &
$ -0.56 \pm  0.08$ &
$ 45.72 \pm  0.33$ &
$-0.88$ \\
$1402+261$ &
$1910$ &
$44.82$ &
$ 7.64$ &
$  0.63$ &
$ -0.40 \pm  0.04$ &
$ 46.07 \pm  0.27$ &
$-1.22$ \\
$1404+226$ &
$ 880$ &
$44.16$ &
$ 6.52$ &
$  0.55$ &
$ -0.26 \pm  0.19$ &
$ 45.21 \pm  0.26$ &
$-2.00$ \\
$1411+442$ &
$2670$ &
$44.45$ &
$ 7.68$ &
$  0.02$ &
$ -1.70 \pm  0.20$ &
$ 45.06 \pm  0.05$ &
$-1.61$ \\
$1415+451$ &
$2620$ &
$44.34$ &
$ 7.59$ &
$ -0.06$ &
$ -0.81 \pm  0.04$ &
$ 45.60 \pm  0.39$ &
$-1.00$ \\
$1416-129$ &
$6110$ &
$44.94$ &
$ 8.74$ &
$ -0.21$ &
$ -0.49 \pm  0.26$ &
$ 45.82 \pm  0.23$ &
$-0.63$ \\
$1425+267$ &
$9410$ &
$45.55$ &
$ 9.53$ &
$  0.07$ &
$ -0.63 \pm  0.07$ &
$ 46.35 \pm  0.20$ &
$-0.38$ \\
$1426+015$ &
$6820$ &
$44.71$ &
$ 8.67$ &
$ -0.49$ &
$ -0.56 \pm  0.07$ &
$ 45.84 \pm  0.24$ &
$-0.32$ \\
$1427+480$ &
$2540$ &
$44.69$ &
$ 7.80$ &
$  0.27$ &
$ -0.75 \pm  0.06$ &
$ 45.64 \pm  0.14$ &
$-1.29$ \\
$1435-067$ &
$3180$ &
$44.90$ &
$ 8.14$ &
$  0.26$ &
$ -1.69 \pm  0.37$ &
$ 45.60 \pm  0.05$ &
$-1.32$ \\
$1440+356$ &
$1450$ &
$44.37$ &
$ 7.09$ &
$  0.43$ &
$ -0.18 \pm  0.06$ &
$ 45.62 \pm  0.29$ &
$-1.47$ \\
$1444+407$ &
$2480$ &
$45.11$ &
$ 8.07$ &
$  0.66$ &
$ -1.03 \pm  0.03$ &
$ 46.28 \pm  0.28$ &
$-1.04$ \\
$1501+106$ &
$5470$ &
$44.18$ &
$ 8.11$ &
$ -0.79$ &
$ -1.26 \pm  0.05$ &
$ 44.90 \pm  0.09$ &
$-0.97$ \\
$1512+370$ &
$6810$ &
$45.48$ &
$ 9.20$ &
$  0.20$ &
$  0.28 \pm  0.05$ &
$ 47.11 \pm  0.50$ &
$ 0.26$ \\
$1519+226$ &
$2220$ &
$44.45$ &
$ 7.52$ &
$  0.18$ &
$  0.67 \pm  0.78$ &
$ 45.98 \pm  0.41$ &
$-0.86$ \\
$1534+580$ &
$5340$ &
$43.63$ &
$ 7.71$ &
$ -1.24$ &
$ -0.93 \pm  0.13$ &
$ 44.49 \pm  0.12$ &
$-0.92$ \\
$1535+547$ &
$1480$ &
$43.90$ &
$ 6.78$ &
$ -0.01$ &
$ -2.59 \pm  0.17$ &
$ 44.34 \pm  0.02$ &
$-2.30$ \\
$1543+489$ &
$1560$ &
$45.27$ &
$ 7.78$ &
$  1.18$ &
$ -2.27 \pm  0.06$ &
$ 46.43 \pm  0.25$ &
$-1.41$ \\
$1545+210$ &
$7030$ &
$45.29$ &
$ 9.10$ &
$  0.01$ &
$ -0.93 \pm  0.10$ &
$ 46.14 \pm  0.13$ &
$-0.53$ \\
$1552+085$ &
$1430$ &
$44.50$ &
$ 7.17$ &
$  0.56$ &
$ -3.29 \pm  0.97$ &
$ 45.04 \pm  0.01$ &
$-2.18$ \\
$1612+261$ &
$2520$ &
$44.54$ &
$ 7.69$ &
$  0.15$ &
$ -1.19 \pm  0.10$ &
$ 45.38 \pm  0.10$ &
$-1.42$ \\
$1613+658$ &
$8450$ &
$44.75$ &
$ 8.89$ &
$ -0.59$ &
$ -0.35 \pm  0.08$ &
$ 45.89 \pm  0.11$ &
$-0.17$ \\
$1617+175$ &
$5330$ &
$44.63$ &
$ 8.40$ &
$ -0.38$ &
$ -0.89 \pm  0.32$ &
$ 45.44 \pm  0.14$ &
$-0.84$ \\
$1626+554$ &
$4490$ &
$44.46$ &
$ 8.13$ &
$ -0.40$ &
$ -0.61 \pm  0.08$ &
$ 45.53 \pm  0.13$ &
$-0.73$ \\
$1704+608$ &
$6560$ &
$45.65$ &
$ 9.29$ &
$  0.38$ &
$ -0.69 \pm  0.05$ &
$ 46.67 \pm  0.21$ &
$-0.36$ \\
$2112+059$ &
$3190$ &
$45.92$ &
$ 8.85$ &
$  1.16$ &
$ -2.47 \pm  0.06$ &
$ 46.47 \pm  0.02$ &
$-1.34$ \\
$2130+099$ &
$2330$ &
$44.35$ &
$ 7.49$ &
$  0.05$ &
$ -0.95 \pm  0.11$ &
$ 45.52 \pm  0.32$ &
$-1.19$ \\
$2209+184$ &
$6500$ &
$44.11$ &
$ 8.22$ &
$ -0.98$ &
$  1.26 \pm  0.52$ &
$ 46.02 \pm  0.47$ &
$ 0.34$ \\
$2214+139$ &
$4550$ &
$44.36$ &
$ 8.08$ &
$ -0.50$ &
$ -0.40 \pm  0.14$ &
$ 45.15 \pm  0.08$ &
$-1.01$ \\
$2251+113$ &
$4160$ &
$45.60$ &
$ 8.86$ &
$  0.66$ &
$ -3.26 \pm  0.06$ &
$ 46.13 \pm  0.01$ &
$-1.18$ \\
$2304+042$ &
$6500$ &
$43.67$ &
$ 7.91$ &
$ -1.35$ &
$ -1.91 \pm  0.84$ &
$ 44.49 \pm  0.03$ &
$-0.81$ \\
$2308+098$ &
$7970$ &
$45.62$ &
$ 9.43$ &
$  0.22$ &
$ -0.38 \pm  0.04$ &
$ 46.61 \pm  0.22$ &
$-0.27$ \\

\enddata

\tablenotetext{a}{The FWHM of the H$\beta$ line in units of $\rm km \; s^{-1}$.}
\tablenotetext{b}{$L_{\rm opt}$ and $L_{\rm bol}$ measured in units of $\rm erg \; cm^{-2} \; s^{-1}$.}
\tablenotetext{c}{$\Mbh$ measured in units of $\Msun$.}
\tablenotetext{d}{$\Mdot$ measured in units of $\Msun \; \rm yr^{-1}$.}
\tablecomments{
Summary of $\Mdot$ and $\eta$ derived using $M_{\rm BLR}$ estimates. 
We report $\Mdot$ and $\Mbh$ without error as systematic uncertainties
in the estimation methods dominate the statistical uncertainty in the
input data.  These uncertainties are discussed further in the text.
For brevity, we do not report the uncertainties in $\log \eta$
because they are identical to the uncertainties in $\log L_{\rm bol}$.}

\end{deluxetable}

\begin{deluxetable}{lccccc}
\tablecolumns{6}
\tablewidth{0pt}
\tablehead{
\colhead{Object} &
\colhead{$\log M_{\sigma}$\tablenotemark{a}} &
\colhead{$\log\Mdot$\tablenotemark{b}} &
\colhead{$\log \eta$}
}
\startdata
$0007$ $+106$ &
$ 8.14$ &
$ -0.26$ &
$-0.88 \pm  0.17$ \\
$0050$ $+124$ &
$ 8.02$ &
$ -0.36$ &
$-1.18 \pm  0.04$ \\
$1119$ $+120$ &
$ 7.76$ &
$ -0.73$ &
$-0.74 \pm  0.34$ \\
$1126$ $-041$ &
$ 8.08$ &
$ -0.75$ &
$-0.75 \pm  0.28$ \\
$1229$ $+204$ &
$ 7.76$ &
$ -0.38$ &
$-1.22 \pm  0.04$ \\
$1302$ $-102$ &
$ 9.09$ &
$  0.62$ &
$-0.76 \pm  0.12$ \\
$1309$ $+355$ &
$ 8.42$ &
$  0.08$ &
$-1.10 \pm  0.05$ \\
$1404$ $+226$ &
$ 8.43$ &
$ -1.07$ &
$-0.38 \pm  0.26$ \\
$1426$ $+015$ &
$ 7.99$ &
$  0.11$ &
$-0.92 \pm  0.24$ \\
$1617$ $+175$ &
$ 7.97$ &
$  0.01$ &
$-1.23 \pm  0.14$ \\
$2130$ $+099$ &
$ 7.87$ &
$ -0.31$ &
$-0.82 \pm  0.32$ \\
$2214$ $+139$ &
$ 7.70$ &
$ -0.14$ &
$-1.37 \pm  0.08$ \\
$1444$ $+407$ &
$ 8.71$ &
$  0.06$ &
$-0.44 \pm  0.28$ \\

\enddata
\tablenotetext{a}{$\Mbh$ measured in units of $\Msun$.}
\tablenotetext{b}{$\Mdot$ measured in units of $\Msun \; \rm yr^{-1}$.}
\tablecomments{
Summary of $\Mdot$ and $\eta$ derived using $M_\sigma$ estimates. 
We report $\Mdot$ and $\Mbh$ without error as systematic uncertainties
in the estimation methods dominate the statistical uncertainty in the
input data.  These uncertainties are discussed further in the text.}

\end{deluxetable}

\end{document}